\documentclass[fleqn,usenatbib,usedcolumn]{mnras}
\usepackage[british]{babel}             
\let\la=\lesssim 
\usepackage[T1]{fontenc}                
\usepackage{graphicx}                   
\usepackage[usenames,dvipsnames]{color}

\hypersetup{pdfauthor={I. Heywood},
               pdftitle={A deep / wide 1--2 GHz snapshot survey of SDSS Stripe 82 using the Karl G.~Jansky Very Large Array in a compact hybrid configuration},
               pdfkeywords={observations: galaxies -- general -- astronomical data bases: surveys},
               bookmarksnumbered=true}

\volume{{\rm in press}}   

\newcommand{\trm}{\textrm}
\newcommand{\gtsim}{\mbox{{\raisebox{-0.4ex}{$\stackrel{>}{{\scriptstyle\sim}}$}}}}

\usepackage{txfonts}
	
\title[A VLA snapshot survey of Stripe 82 at 1--2 GHz]{A deep / wide 1--2 GHz snapshot survey of SDSS Stripe 82 using the Karl G.~Jansky Very Large Array in a compact hybrid configuration}

\author[Heywood et al.]
{\parbox{\textwidth}{
\begin{flushleft}
I.~Heywood$^{1,2}$\thanks{Contact author: {\tt ian.heywood@csiro.au}}, 
M.~J.~Jarvis$^{3,4}$,
A.~J.~Baker$^{5}$,
K.~W.~Bannister$^{1}$, 
C.~S.~Carvalho$^{6,7}$, 
M.~Hardcastle$^{8}$,
M.~Hilton$^{9}$,
K.~Moodley$^{9}$,
O.~M.~Smirnov$^{2,10}$,
D.~J.~B.~Smith$^{8}$, 
S.~V.~White$^{11,3}$, 
E.~J.~Wollack$^{12}$\\
\end{flushleft}
}
\footnotesize
\\
$^{1}$CSIRO Astronomy and Space Science, P.O. Box 76, Epping, NSW 1710, Australia\\
$^{2}$Department of Physics and Electronics, Rhodes University, PO Box 94, Grahamstown, 6140, South Africa\\
$^{3}$Astrophysics, Department of Physics, Keble Road, Oxford, OX1 3RH\\ 
$^{4}$Physics Department, University of the Western Cape, Private Bag X17, Bellville 7535, South Africa\\
$^{5}$Department of Physics and Astronomy, Rutgers, The State University of New Jersey, 136 Frelinghuysen Road,
Piscataway, NJ 08854-8019, USA\\
$^{6}$Institute of Astrophysics and Space Sciences, University of Lisbon, Tapada da Ajuda, 1349-018 Lisboa, Portugal\\
$^{7}$Research Center for Astronomy and Applied Mathematics, Academy of Athens, Soranou Efessiou 4, 11-527, Athens, Greece\\
$^{8}$Centre for Astrophysics, Science and Technology Research Institute, University of Hertfordshire, Hatfield, Herts AL10 9AB, UK\\
$^{9}$Astrophysics and Cosmology Research Unit, School of Mathematics, Statistics \& Computer Science, University of KwaZulu-Natal, Durban 4041, South Africa\\
$^{10}$SKA South Africa, 3rd Floor, The Park, Park Road, Pinelands, 7405, South Africa\\
$^{11}$International Centre for Radio Astronomy Research (ICRAR), Curtin University,  Bentley, WA 6102, Australia	\\
$^{12}$Observational Cosmology Laboratory, Code 665, NASA, Goddard Space Flight Center, Greenbelt, MD, USA}

\date{Last updated 20XX Jan 01; in original form 20XX Jan 01}

\pubyear{2016}

\begin{document}
\label{firstpage}
\pagerange{\pageref{firstpage}--\pageref{lastpage}}
\maketitle

\begin{abstract}

We have used the Karl G.~Jansky Very Large Array to image $\sim$100 deg$^{2}$ of SDSS Stripe 82 at 1--2 GHz. The survey consists of 1,026 snapshot observations of 2.5 minutes duration, using the hybrid CnB configuration. The survey has good sensitivity to diffuse, low surface brightness structures and extended radio emission, making it highly synergistic with existing 1.4 GHz radio observations of the region. The principal data products are continuum images, with 16~$\times$~10 arcsecond resolution, and a catalogue containing 11,782 point and Gaussian components resulting from fits to the thresholded Stokes-I brightness distribution, forming approximately 8,948 unique radio sources. The typical effective 1$\sigma$ noise level is 88 $\mu$Jy beam$^{-1}$. Spectral index estimates are included, as derived from the 1~GHz of instantaneous bandwidth. Astrometric and photometric accuracy are in excellent agreement with existing narrowband observations. A large-scale simulation is used to investigate clean bias, which we extend into the spectral domain. Clean bias remains an issue for snapshot surveys with the VLA, affecting our total intensity measurements at the $\sim$1$\sigma$ level. Statistical spectral index measurements are in good agreement with existing measurements derived from matching separate surveys at two frequencies. At flux densities below $\sim$35$\sigma$ the median in-band spectral index measurements begin to exhibit a bias towards flatness that is dependent on both flux density and the intrinsic spectral index. In-band spectral curvature measurements are likely to be unreliable for all but the very brightest components. Image products and catalogues are publicly available via an FTP server.

\end{abstract}





\section{Introduction}
\label{sec:intro}

Over the past decade it has become clear that in order to understand the formation and evolution of galaxies we need to account for the various feedback processes that are able to both stimulate and truncate star formation. These are generally split into three forms of feedback. In the less massive galaxies supernovae provide the necessary energy to drive the gas from these relatively shallow potential wells, providing negative feedback that is able to truncate the star formation \citep[e.g.][]{Efstathiou2000,DallaVecchia2008}. In more massive galaxies, supernovae do not provide enough energy to drive the gas out of more massive haloes, and therefore an alternative mechanism is required. Negative feedback from active galactic nuclei (AGN) was introduced to models and simulations of galaxy formation in order to halt the star formation in more massive galaxies \citep[e.g.][]{Bower2006,Croton2006,Sijacki2007}. Since the initial introduction of feedback from AGN in simulations, further progress has been made in differentiating between radiative (or cold-mode) feedback and the mechanical (or hot-mode) feedback that results principally from the passage of radio jets and lobes \citep[e.g.][]{Hardcastle2007,BestHeckman2012,Fabian2012,Fernandes2015}. However, it is clear that some level of positive feedback also occurs as a result of AGN activity: radio jets sometimes appear to stimulate star formation as they pass through the interstellar and intergalactic medium \citep[e.g.][]{Croft2006,SilkNusser2010,Silk2013,Kalfountzou2012,Kalfountzou2014}.

To differentiate between these forms of feedback, and to trace their evolution, requires observations that are both deep enough to detect star formation and low-luminosity radio jets in galaxies, and wide enough to obtain the number statistics with which to trace the evolution of the rarer bright sources. Radio continuum observations provide a unique opportunity to study both the black-hole accretion activity in the Universe \citep[e.g.][]{JarvisRawlings2000,Smolcic2009,Rigby2011,McAlpine2013,Best2014}, and the star-formation history of the Universe \citep[e.g.][]{Seymour2008,Smolcic2009sf,Jarvis2015a}, at a wavelength that is free from selection biases due to dust obscuration. 

Additionally, it has been suggested that the radio continuum surveys envisaged to be carried out with the new generation of radio telescopes will provide a complementary technique to measure the large-scale structure of the Universe, but to much higher redshifts and volumes than is feasible with optical surveys \citep[e.g.][]{Raccanelli2012,Camera2012,Ferramacho2014,Jarvis2015b}. Although these are predominantly the remit of the very deep all-sky surveys that will be carried out with the Square Kilometre Array \citep{Dewdney2013}, much progress in this area can be made now by obtaining a greater understanding of the properties of radio sources over fields targeted due to their extensive multi-wavelength data.

Various deep-field surveys have been conducted with the (Karl G.~Jansky) Very Large Array \citep[VLA; e.g.][]{Bondi2003,Simpson2006,Schinnerer2007,Miller2013,Heywood2013}, the Westerbork Synthesis Radio Telescope \citep[e.g.][]{devries2002}, the Giant Metrewave Radio Telescope \citep[e.g.][]{Garn2008} and the Australia Telescope Compact Array \citep[e.g.][]{Norris2006,Middelberg2008}. All of these cover areas of the order of 1-10~deg$^2$, which, although very good for detecting the bulk of the typical radio sources that lie close to the flux-density limit, miss the rarer populations, such as the most luminous starbursts and powerful AGN. The volume surveyed also limits the investigation of the environmental influence in triggering or truncating AGN and/or star-formation activity at relatively low redshift \citep[e.g.][]{Peng2010,Burton2013,Karouzos2014,Sabater2014}. For this reason, projects such as the Galaxy and Mass Assembly \citep[GAMA; ][]{Driver2011,Hopkins2013} survey, along with multi-wavelength data at radio \citep{Mauch2013} and far-infrared \citep{Eales2010} wavelengths, provide an important tier in the general wedding-cake structure of extragalactic surveys.

The SDSS Stripe~82 region is quickly becoming the field of choice for surveys covering $\ge 100$~square degrees. This is due to the wealth of multi-wavelength data that has been accrued following the repeat observations in $ugriz$ as part of the SDSS Survey, reaching $g\sim 24.5$ \citep{Jiang2014,Annis2014}. Additional multi-wavelength data include near-infrared imaging ($YJHK$) as part of the UKIRT Infrared Deep Sky Survey \citep[UKIDSS; ][]{Lawrence2007} and more recently the VISTA Hemisphere Survey \citep[VHS;][]{VHS} to $K_{\rm AB}\sim 20$,  far-infrared imaging data from the {\em Herschel Space Observatory} as part of the {\em Herschel} Multi-tiered Extragalactic Survey \citep[HerMES;][]{Oliver2012} and {\em Herschel} Stripe 82 Survey \citep[HerS;][]{Viero2014}, and at millimetre-wavelength imaging with the Atacama Cosmology Telescope \citep[ACT;][]{Hincks2010}. A survey covering 92~square degrees within Stripe 82 to a median flux-density limit of 52~$\mu$Jy beam$^{-1}$ at 1.4 GHz has already been carried out with the pre-upgrade VLA in its most extended A-array configuration \citep{Hodge2011}. This dataset provides excellent resolution ($\sim$1.5~arcsec) imaging with sufficient positional accuracy to allow cross-matching to the wealth of ancillary data. However, the A-array data also resolves out the more-diffuse emission from lobes and any possible extended emission from star-forming galaxies. They do not therefore obtain an accurate measure of the total flux density from these systems. Furthermore, objects identified as single point sources in the \citet{Hodge2011} data may in fact turn out to be components of a single extended source.

We have used the Karl G.~Jansky Very Large Array in a compact, hybrid CnB configuration to target the area covered by the observations of \citet{Hodge2011}. This configuration results in excellent sensitivity to diffuse and low surface brightness radio structures, with short baselines sensitive to structures up to $\sim$16 arcmin in extent, making the observations highly complementary to the existing data. The use of 1 GHz of uninterrupted bandwidth means that the new observations can cover the survey area via a series of short snapshot observations and reach depths comparable to those of the existing observations. 

\section{Observations and data processing}
\label{sec:selfcal}

\begin{figure*}
\begin{center}
\setlength{\unitlength}{1cm}
\includegraphics[width = \textwidth]{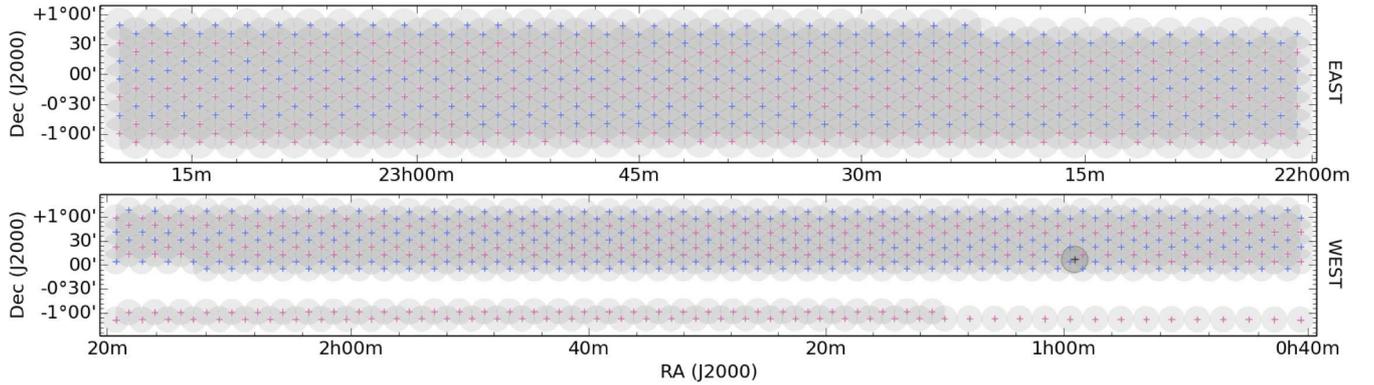}
\caption{Coverage of the observed pointings of the survey described in this paper for the eastern (upper) and western (lower) area. The circles show the approximate size of the half-power point of the VLA primary beam at 1.5~GHz. The set of pointings contained within an individual Scheduling Block (SB) are coloured alternately with blue and pink markers. The eastern patch contains 522 pointings. The extra pointing in the western area corresponds to an image derived from the visits to the 2.9~Jy phase calibrator J0059+0006, added to the mosaic to improve the image quality around this source, bringing the total number of pointings contributing to the final western mosaic to 505. Note the discontinuity in the western patch due to one of the Scheduling Blocks not being executed.}
\label{fig:pointings}  
\end{center}
\end{figure*}

The data\footnote{Project code: 13B-272} were taken with the array in the CnB\footnote{This is a hybrid configuration in which the east and west arms of the array are spaced as they are in C configuration and the northern arm of the array is spaced according to the more extended B configuration.} configuration. Standard wide-band mode was employed with the correlator splitting the 1--2 GHz of frequency coverage into 16 spectral windows (SPWs) with 64~$\times$~1~MHz channels each, and an integration time per visibility point of 3 seconds. The programme was divided into sixteen scheduling blocks (SBs) of 4.5 hours duration, each of which was self-contained with both target scans and appropriate calibration observations. The SBs were scheduled dynamically by the observatory. A total of 1368 target pointings were scheduled, 608 and 760 in the eastern and western regions respectively, coincident with the two eastern and western areas of the existing \citet{Hodge2011} data. The observations were designed to follow a standard hexagonal mosaic pattern, with pointings lying at the half-power point of their neighbour in right ascension, such that they are critically sampling the sky at the centre of the frequency band. The total integration time per pointing was approximately 2.5 minutes. 3C48 was observed once per SB for flux calibration, and the strong calibrator J2225-0457 was observed for bandpass calibration. A suitable phase calibrator source within the survey area was visited approximately once per hour. When the array moved from its CnB configuration, 12 of the 16 SBs had been observed resulting in 522 and 504 observed target pointings in the eastern and western regions respectively, and a somewhat unfortunate discontinuity in the coverage of the western patch. The pointing positions and the total sky coverage are shown in Figure \ref{fig:pointings}, with the alternating colours used to demarcate different SBs.

Initial processing of the data was performed using the NRAO VLA pipeline\footnote{{\tt \tiny \href{https://science.nrao.edu/facilities/vla/data-processing/pipeline}{https://science.nrao.edu/facilities/vla/data-processing/pipeline}}}. This is a set of {\tt CASA}\footnote{\href{http://casa.nrao.edu/}{http://casa.nrao.edu/}} \citep{McMullin2007} scripts that were used to apply Hanning smoothing and make a first pass of flagging bad or unnecessary data. The flagging steps include removal of data from shadowed antennas, the initial few integration points of a scan where not all antennas may be on source and visibility amplitudes that are exactly zero, together with automatic radio frequency interference (RFI) excision using using the sliding time median filtering approach of the {\tt rflag} algorithm. Following these steps the pipeline performs delay and bandpass corrections. Antenna-based complex gain corrections were derived from the calibrator scans and interpolated over the target fields. No post-hoc time or frequency averaging of the 3~s / 1~MHz native resolutions was performed. Retaining these time and frequency resolutions for this array configuration means that time and bandwidth smearing effects are of no concern, the latter mitigated by the imaging method described below. During the observations, visits to a strong source for phase-calibration  were scheduled somewhat infrequently in order to minimise overheads, on the grounds that there would be enough flux in any given field to further correct complex gain errors via self-calibration. 

Following execution of the NRAO pipeline, each of the target pointings (typically $\sim$90 per SB) was split into a single Measurement Set and a secondary pass of automatic flagging was performed on the calibrated data using the {\tt rflag} algorithm within the {\tt CASA} {\tt flagdata} task. Each pointing was then subjected to a self-calibration procedure. 

An estimate of the thermal noise for each pointing was made by measuring the standard deviation of a 4096~$\times$~4096 1.5 arcsec-pixel Stokes-V image, and the dimensions of the fitted restoring beam specific to that pointing were determined. Total intensity images were then made, spanning 4096~$\times$~4096 pixels (again with a 1.5 arcsec pixel scale), encompassing the first sidelobe of the VLA primary beam pattern at the lowest frequency. Multi-term, multi-frequency synthesis (MT-MFS) imaging \citep{Rau2011} was used, as implemented in {\tt CASA}'s {\tt clean} task. Briefly, the clean components are modelled in the frequency dimension by a Taylor polynomial 
\begin{equation}
I(\nu)~=~\sum_{t}I_{t}\left(\frac{\nu - \nu_{0}}{\nu_{0}}\right)^{t},
\end{equation}
where $I(\nu)$ is the component brightness as a function of frequency, $I_{t}$ is the image of Taylor polynomial coefficient $t$, and $\nu_{0}$ is the reference frequency about which the expansion is performed, set in this case to the band centre at 1.5 GHz. A typical choice for the spectral model, motivated by the frequency behaviour of astrophysical synchrotron emission, is a power law with a curvature term of the form
\begin{equation}
I(\nu)~=~I_{\nu_{0}}\left(\frac{\nu}{\nu_{0}}\right)^{\alpha+\beta \mathrm{log}(\nu/\nu_{0})}
\label{eq:specmodel}
\end{equation}
where $I_{\nu_{0}}$ is the sky brightness at the reference frequency, $\alpha$ is the power law index (the spectral index\footnote{We adopt the convention $S$~$\propto$~$\nu^{\alpha}$ where $S$ is the flux density, $\nu$ is the frequency and $\alpha$ is the spectral index.}) and $\beta$ is the curvature term. A third-order ($t$~=~0,1,2) Taylor expansion of Equation \ref{eq:specmodel} then returns the total intensity image $I_{\nu_{0}}$~=~$I_{0}$, and allows $\alpha$ and $\beta$ to be determined from linear combinations of the higher-order Taylor coefficient images according to
\begin{equation}
\alpha~=~\frac{I_{0}}{I_{1}}
\label{eq:alpha}
\end{equation}
and
\begin{equation}
\beta~=~\frac{I_{2}}{I_{0}}~-~\frac{\alpha(\alpha - 1)}{2}.
\label{eq:beta}
\end{equation}
A third order Taylor series was employed during deconvolution as, based on tests of a limited subset of the data, it delivered improved dynamic range in the Stokes-I images, particularly for pointings that included off-axis sources with brightnesses in the $\sim$10--100 mJy beam$^{-1}$ range. The full frequency resolution gridding technique that the algorithm adopts renders the effects of bandwidth smearing negligible.

The low declination of Stripe 82, the extended northern arm of the array in CnB configuration and the short snapshot pointings result in a point spread function (PSF) with an elongated main lobe and high sidelobe levels. The use of \citet{Briggs1995} weighting can temper these effects by weighting down the data in the inner part of the $u$,$v$ plane: in broad terms, higher angular resolution and lower sidelobe levels can be achieved at the expense of reduced sensitivity. Down-weighting the short spacings also affects the sensitivity to extended structure, however deconvolution is made significantly easier and more reliable with the application of a weighting function that moves further from natural weighting to uniform weighting. After some trial runs on a limited subset of the data we opted for a robust parameter of 0.3, reasoning that the unique combination of depth and sensitivity to extended emission that these observations have should be retained given the properties of the existing radio observations of Stripe 82. 

Deconvolution of a snapshot VLA survey must be done with extreme care. The suppression of source brightnesses in an improperly cleaned image (an effect known as clean or snapshot bias, discussed further in Section \ref{sec:photometry}) is a well known but poorly understood phenomenon. The use of masks to constrain the regions that are subjected to cleaning is known to reduce the effect, as is terminating the cleaning cycle before spurious sidelobe features or noise peaks are cleaned. Thus for each pointing a cleaning mask was generated by convolving the positions of sources from the Faint Images of the Radio Sky at Twenty-cm \citep{Becker1995} catalogue with the Gaussian restoring beam appropriate to that pointing, and converting this into a Boolean mask above the 5$\sigma$ level, where $\sigma$ was the value determined from the initial Stokes-V image.

The images were deconvolved using this mask, terminating the process after 8,000 iterations or when the peak residual dropped below 5$\sigma$. We generated model visibilities (including spectral behaviour) from the Taylor term images that resulted from this process, relying on the use of the cleaning mask to prevent spurious sidelobe features from entering the model. Phase-only gain corrections were then generated for each of the 16 spectral windows using a single solution interval for the 2.5 minute scan. Data that did not meet minimum signal-to-noise or number of baselines criteria when forming a solution were flagged.

The data were then re-imaged using the FIRST-based cleaning mask. For the final images, a second pass of deconvolution was applied with the mask removed, with deconvolution terminating after 2000 iterations or when the peak in the residual image reached 3$\sigma$, whichever was sooner. These final images were restored using a 16 arcsec~$\times$~10 arcsec Gaussian with a position angle (PA) of 90$^{\circ}$ east of north. This is slightly broader than the generally-achievable angular resolution, however it imparts a desirable level of uniformity to the survey products, and it encompasses the variation in the fitted restoring beams that is caused by the dynamic scheduling of the observations. Mean values plus/minus one standard deviation of the distribution of per pointing major axes, minor axes and PAs for the survey are (13.3~$\pm$~2.1) arcsec, (7.5~$\pm$~0.3) arcsec and (88~$\pm$~15) deg respectively.

The {\tt widebandpbcor} task was used to correct for primary beam effects in the image domain. This task works by making a spectral cube of a model of the main lobe of the VLA's primary beam, in this case with one frequency plane for each of the 16 SPWs. This cube represents the directional sensitivity of the array as a function of frequency. The frequency cube is then transformed into 2D images for each Taylor term $t$, in this case for $t$~=~[0,1,2], which are then used to correct the Taylor term images of the sky brightness distribution. The $t$~=~0 image is practically equivalent to the usual narrow-band correction method, i.e. it is divided by a total intensity beam model computed at the band centre. Examining the difference between the {\tt widebandpbcor} $t$~=~0 image and the effective sensitivity image produced by the {\tt clean} task shows deviations of around 1--2\% across the field. Accurate primary beam-corrected flux densities are recovered, aided by the spacing of the pointings and the weighting method used when forming the mosaic (see below). The higher-order Taylor term images should in principle now contain intrinsic spectral index and curvature estimates, with the spectral corruption effects of the primary beam removed. Once the wide band primary beam correction has been applied to the three Taylor term images the spectral index ($\alpha$) and spectral curvature maps ($\beta$) are recomputed as per Equations \ref{eq:alpha} and \ref{eq:beta}. All primary beam corrected images were cut beyond the radius where the normalized, total-intensity primary beam gain as computed at the band centre exceeded 0.4.

At this point the total intensity images were examined to identify pointings for which the automatic calibration procedure had been inadequate. Such datasets invariably fell into one of the following categories: (i) a very strong source inside the imaged area; (ii) a strong source outside of the imaged area, the sidelobes of which were encroaching into the main image; (iii) morphologically complex sources, or very faint extended sources, for which the FIRST-based cleaning mask was inadequate; (iv) residual RFI that had been missed by the automatic flagging.

Case (iv) was remedied by manually editing the visibility data and reprocessing the Measurement Set using the steps described above. Cases (ii) and (iii) were addressed by the addition of outlier fields at the position of the confusing sources, and by setting manual cleaning masks. This was done on the post-NRAO pipelined data, discarding the automatically generated phase-self-calibration solutions. Having generated a model via this process that was certainly more complete than the model used for the automated self-calibration procedure, we executed three iterations of imaging and self-calibration, two passes of phase only self-calibration (with 30 s and 6 s solution intervals), and one pass of amplitude and phase self-calibration with a single correction applied to the whole 2.5 minute scan. Solutions were generated in each of the 16 SPWs. This extended self-calibration procedure was also used to improve case (i) pointings.  Primary beam correction was then applied as described above. Of the 1,026 pointings in the survey, 202 (20 percent) of them required manual re-processing, and thus a smaller fraction than this were subjected to amplitude self-calibration. This subset also includes fields for which the automated calibration procedure was sufficient, but had significantly extended sources in the survey which only required a deeper (interactive) deconvolution process than our automated procedure afforded.

Improper amplitude self-calibration can bias the flux density measurements in the resulting map. This is due to the risk of the contribution to the visibility function made by the faintest sources which are typically absent from the sky model being subsumed into the gain solutions. The end result of this is suppression of the unmodelled sources, as well as more subtle, counterintuitive effects, e.g. \citet{grobler2014}, see also \citet{franzen2015} and \citet{heywood2016}. We do not expect to suffer significantly from such self-cal biases, as (i) the model should be largely complete, being formed via interactive, constrained deconvolution which was terminated when the residuals contained only noise, or noise plus artefacts; (ii) our solution interval was conservative, with only a single amplitude and phase correction being applied for the entire scan; and (iii) the VLA is robust against such biases due to the self calibration problem being well constrained, with 27 complex gain terms being derived from 351 equations.

Pointings that were processed manually were re-imaged using their individual clean masks and primary beam corrected following the procedures used in the automated case. Once all the final images were in place, the {\tt Montage}\footnote{{\tt \href{http://montage.ipac.caltech.edu/}{http://montage.ipac.caltech.edu/}}} software was used to construct linear mosaics in total intensity, $\alpha$, $\beta$, and $\alpha$-error. For all mosaics, each pointing is weighted by the assumed variance in the total intensity map (i.e.~the square of the assumed Stokes-I primary beam).

\section{Data products}
\label{sec:results}

The principal products from this snapshot survey are wide area mosaics for the east and west regions, formed directly (Stokes-I, $\alpha$, $\beta$, $\alpha$-error) or derivatively (model, residual and sensitivity) from the individual pointing images as described in Section \ref{sec:selfcal}, and the corresponding lists of components derived from them using automatic source-finding software.

\subsection{Radio mosaics}
\label{sec:mosaic}

\begin{figure*}
\begin{center}
\setlength{\unitlength}{1cm}
\includegraphics[width = \textwidth]{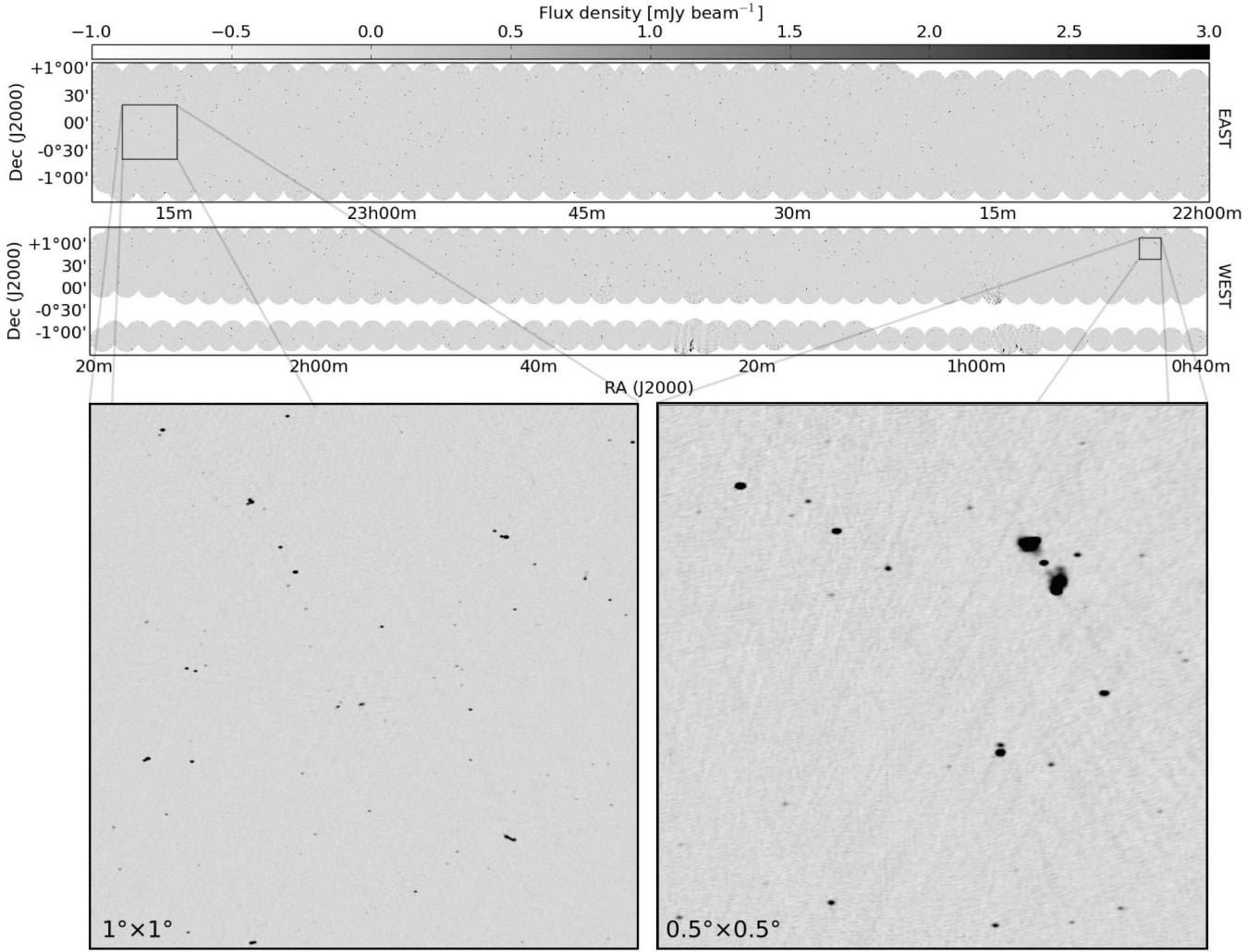}
\caption{Total intensity mosaics showing the full extent of the survey for the eastern and western regions (upper two panels). The pixel scale is linear, is the same for all panels in the image, and runs from -1.0 to 3.0 mJy beam$^{-1}$ as indicated by the colour bar at the top of the figure. The square markers in the eastern and western regions are 1$^{\circ}$~$\times$~1$^{\circ}$ and 0.5$^{\circ}$~$\times$~0.5$^{\circ}$ in extent respectively. A closer view of these regions is shown in the lower panels of the figure. The striations most evident in the western close-up region are associated with the PSF of the observations, and remain in the image due to imperfect deconvolution.}
\label{fig:maps} 
\end{center}
\end{figure*}

Stokes-I mosaics of the east and west regions are shown in Figure \ref{fig:maps}. As each mosaic contains approximately a third of a billion pixels only the brightest sources are obvious in the upper panels showing the full area. The discontinuity in the western field due to the unobserved SBs is apparent as the horizontal gap spanning the width of the field. Representative zooms of a 1$^{\circ}$~$\times$~1$^{\circ}$ region in the eastern field and a 0.5$^{\circ}$~$\times$~0.5$^{\circ}$ region in the western field can be seen in the lower panels. The pixel brightness scale runs from $-$1 to 3~mJy~beam$^{-1}$ for all components of Figure \ref{fig:maps}, as indicated by the colour bar at the top. The image products are examined further in Section \ref{sec:imgcomp}.

\subsection{Component catalogue}
\label{sec:catalogue}

\begin{figure*}
\begin{center}
\setlength{\unitlength}{1cm}
\includegraphics[width = \textwidth]{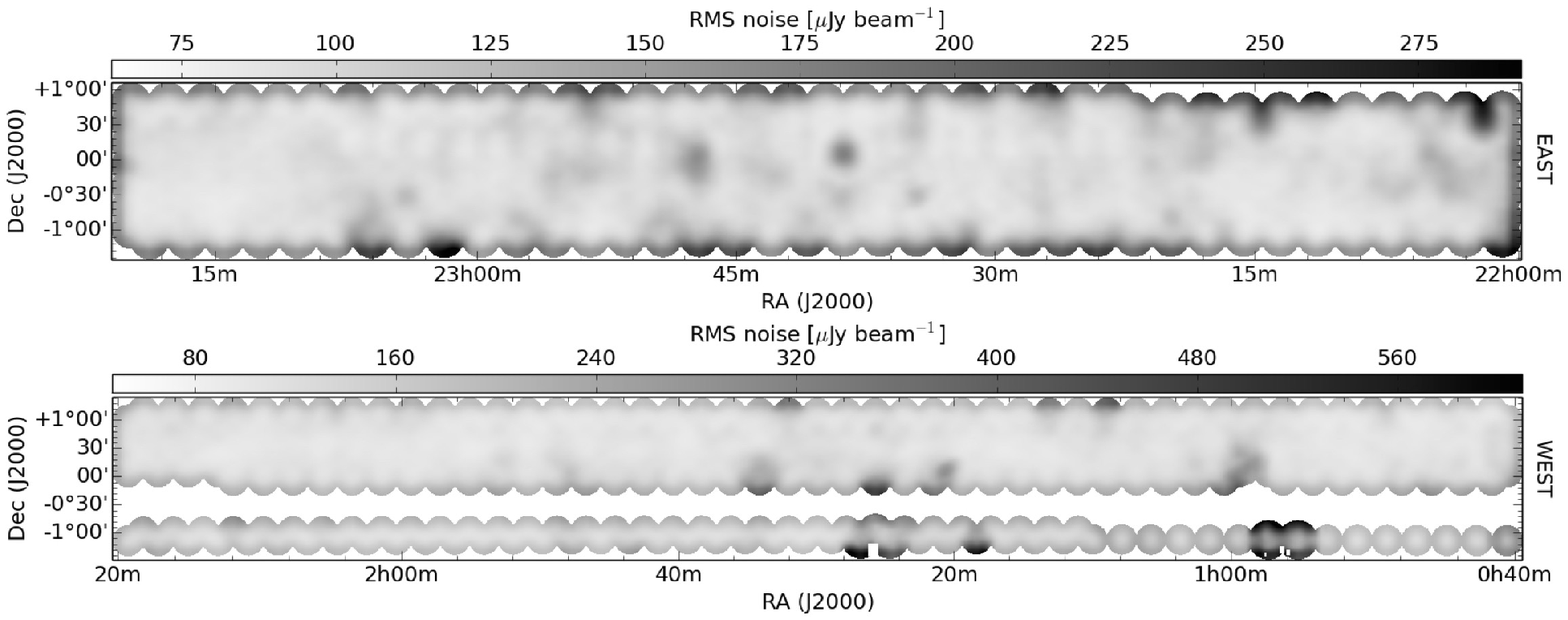}
\caption{Variations in the image noise across the east (upper) and west (lower) mosaic areas shown as a greyscale image. The corresponding values are shown in the colour bars above each panel. The noise is higher at the edge of the mosaic area due to the primary beam correction also raising the image noise, and also around strong or very extended sources where calibration deficiencies or incomplete deconvolution leave PSF-like residuals that effectively raise the noise floor. Components in the catalogue (Section \ref{sec:catalogue}) have the local RMS at their position listed based on the measurement from the above image. The masked rectangular regions visible at the southern edge of the western field (at approximate RAs of 0h57m and 01h25m) correspond to three very extended sources that {\tt PyBDSM} was unable to decompose into a point and Gaussian model (see Section \ref{sec:sources}).}
\label{fig:rms} 
\end{center}
\end{figure*}

The source catalogue for this survey was generated by running the Python Blob Detection and Source Management ({\tt PyBDSM}) source finder \citep{MohanRafferty2015} on each of the wide area radio images presented in Section \ref{sec:mosaic}. A brief overview of the source detection algorithm is as follows. The spatial variation of the image noise was estimated by moving a box across the image in overlapping steps, calculating the root mean square (RMS) of the pixels within the box and interpolating the values measured from each step. The box size was 250 $\times$ 250 pixels, stepped across the image in units of 50 pixels. This procedure provided the source finder with an estimate of the spatial variation of the image noise for detection thresholding purposes. This variation was written to a FITS image in order to provide a map of the sensitivity of the eastern and western fields, and these images are presented in Figure \ref{fig:rms}. 

Image peaks that exceeded 5$\sigma$ where $\sigma$ is the local value of the RMS noise are identified. These detections were then grown into islands, defined as regions where there is contiguous emission above a secondary threshold, in this case 3$\sigma$. The islands were then decomposed into a component list by fitting Gaussian components to them. An attempt is made to measure the intrinsic shape of each component by deconvolving the restoring beam from the component, and then fitting a Gaussian component to the result. If no satisfactory fits are obtained then the component is deemed to be indistinguishable from being point-like.

{\tt PyBDSM} also attempts to identify individual sources within each island via a pairwise examination of the fitted components. The criteria which must be satisfied for components to be part of the same source are (i) the pixels along the line joining the two component centres must all be above the 3$\sigma$ island threshold, and (ii) the length of this line is less than the sum of the full width half maxima of the two components. 
Generating a source catalogue from a radio survey in most cases involves a trade-off between maximising the catalogue completeness and minimising the risk of including of spurious features. Although setting a cutoff at 5$\sigma$ will naturally result in the exclusion of numerous faint sources, it has long been established that below this threshold Eddington bias (Eddington, 1913) will preclude sensible interpretation of the results due to the inclusion of very high numbers of spurious noise features \citep{Condon1974,Hogg2001}.

Around strong sources in particular, incomplete deconvolution and residual calibration errors can leave spurious PSF-like features. The Y-shape of the VLA results in linear structure with six-fold symmetry in the PSF for any observation that is not sufficiently long enough to achieve complete Fourier plane coverage, and for snapshot surveys such as this one these features are very pronounced. As mentioned in Section \ref{sec:selfcal}, even with tapered weighting of the visibilities the PSF exhibits sidelobes at 10--20 percent of the peak level. This exacerbates the error patterns around sources that do not deconvolve properly, and in numerous parts of the mosaic components that do not represent genuine astronomical signals will find their way into the source catalogue. The position of every entry in the component list with a peak flux density exceeding 100 mJy beam$^{-1}$ was therefore examined and any visually identified artefacts were removed. Pointings with known calibration deficiencies (typically around the phase calibrator J0059+0006) were also examined for spurious features. In general the use of a local RMS estimate as opposed to fixed threshold results in a catalogue that is not flux limited but does have high reliability. The spectral indices presented in the catalogue are discussed in detail in Section \ref{sec:alphas}.

The final eastern catalogue contains 5,674 individual point and Gaussian components, fitted to 4,602 islands and grouped into 4,354 sources. The western catalogue contains 6,094 components, 4,869 islands and 4,594 sources.

\begin{table*}
\begin{minipage}{170mm}
\centering
\caption{The first ten components from the eastern catalogue, presented here in order to demonstrate the table structure. Note that the local RMS values here are not typical as the first components in the catalogue are extracted from the edge of the mosaic area. Please refer to the text for a detailed description of each column.}\label{tab:components}
\begin{tabular}{rcccccccccc} \hline
	 &
	\multicolumn{1}{c}{ID} &
	\multicolumn{1}{c}{R.A.} &
	\multicolumn{1}{c}{Decl.} &
	\multicolumn{1}{c}{$\sigma_{\trm{\tiny R.A.}}$} &
	\multicolumn{1}{c}{$\sigma_{\trm{\tiny Decl.}}$} &
	\multicolumn{1}{c}{$S_{\trm{\tiny int}}$} &
	\multicolumn{1}{c}{$\sigma_{S_\trm{\tiny int}}$} &
	\multicolumn{1}{c}{$S_{\trm{\tiny peak}}$} &
	\multicolumn{1}{c}{$\sigma_{S_\trm{\tiny peak}}$} \\
	&
	&
	\multicolumn{1}{c}{[deg]} &
	\multicolumn{1}{c}{[deg]} &
	\multicolumn{1}{c}{[arcsec]} &
	\multicolumn{1}{c}{[arcsec]} &
	\multicolumn{1}{c}{[mJy]} &
	\multicolumn{1}{c}{[mJy]} &
	\multicolumn{1}{c}{[mJy~b$^{-1}$]} &
	\multicolumn{1}{c}{[mJy~b$^{-1}$]} \\
	&
	\multicolumn{1}{c}{(1)} &
	\multicolumn{1}{c}{(2)} &
	\multicolumn{1}{c}{(3)} &
	\multicolumn{1}{c}{(4)} &
	\multicolumn{1}{c}{(5)} &
	\multicolumn{1}{c}{(6)} &
	\multicolumn{1}{c}{(7)} &
	\multicolumn{1}{c}{(8)} &
	\multicolumn{1}{c}{(9)} \\ \hline
\textcolor{Gray}{\emph{1}}& J232038.0+003139& 350.15832&   0.52752&  0.02 &  0.01 &    80.21&     0.30&    73.98&     0.17& \\
\textcolor{Gray}{\emph{2}}&J232026.9-002731&  350.11217&   -0.45875&    0.04&    0.01&    38.78&     0.26&    35.90&     0.15& \\
\textcolor{Gray}{\emph{3}}&J232032.1+002351&  350.13383&    0.39765&    0.06&    0.03&    27.23&     0.29&    24.36&     0.17& \\
\textcolor{Gray}{\emph{4}}&J232049.2+001430&  350.20493&    0.24173&    0.28&    0.14&     7.07&     0.27&     5.25&     0.17& \\
\textcolor{Gray}{\emph{5}}&J232038.9+002718&  350.16226&    0.45522&    0.43&    0.17&     3.23&     0.30&     3.29&     0.16& \\
\textcolor{Gray}{\emph{6}}&J232038.1-005554&  350.15886&   -0.93193&    0.48&    0.20&     3.20&     0.26&     2.82&     0.15& \\
\textcolor{Gray}{\emph{7}}&J232034.3-004254&  350.14293&   -0.71517&    1.49&    0.45&     1.20&     0.24&     1.00&     0.14& \\
\textcolor{Gray}{\emph{8}}&J232031.3-002624&  350.13030&   -0.44017&    1.25&    0.59&     1.10&     0.26&     0.99&     0.15& \\
\textcolor{Gray}{\emph{9}}&J232039.4-001318&  350.16421&   -0.22176&    0.99&    1.72&     0.92&     0.27&     0.75&     0.17& \\
\textcolor{Gray}{\emph{10}}&J232039.5-000037&  350.16453&   -0.01029&    0.84&    0.40&     0.63&     0.34&     1.11&     0.15& \\
 \hline
\end{tabular}
\vspace{2mm}
\begin{tabular}{rcccccccccccc} \hline
	&
	\multicolumn{1}{c}{Local RMS} &
	\multicolumn{1}{c}{$\alpha$} &
	\multicolumn{1}{c}{$\theta_{\mathrm{maj}}$} &
	\multicolumn{1}{c}{$\sigma_{\theta_{\mathrm{maj}}}$} &
	\multicolumn{1}{c}{$\theta_{\mathrm{min}}$} &
	\multicolumn{1}{c}{$\sigma_{\theta_{\mathrm{min}}}$} &
	\multicolumn{1}{c}{PA} &
	\multicolumn{1}{c}{$\sigma_{PA}$} &
	\multicolumn{1}{c}{Resolved}&
	\multicolumn{1}{c}{ID$_{g}$} &
	\multicolumn{1}{c}{ID$_{s}$} &
	\multicolumn{1}{c}{ID$_{i}$} \\
	&
	\multicolumn{1}{c}{[mJy~b$^{-1}$]} &
	\multicolumn{1}{c}{} &
	\multicolumn{1}{c}{[arcsec]} & 
	\multicolumn{1}{c}{[arcsec]} &
	\multicolumn{1}{c}{[arcsec]} &
	\multicolumn{1}{c}{[arcsec]} &
	\multicolumn{1}{c}{[deg]} &
	\multicolumn{1}{c}{[deg]} &
	&
	&
	&\\
	&
	\multicolumn{1}{c}{(10)} &
	\multicolumn{1}{c}{(11)} &
	\multicolumn{1}{c}{(12)} &
	\multicolumn{1}{c}{(13)} &
	\multicolumn{1}{c}{(14)} &
	\multicolumn{1}{c}{(15)} &
	\multicolumn{1}{c}{(16)} &
	\multicolumn{1}{c}{(17)} &
	\multicolumn{1}{c}{(18)} &
	\multicolumn{1}{c}{(19)} &
	\multicolumn{1}{c}{(20)} &
	\multicolumn{1}{c}{(21)}\\ \hline
\textcolor{Gray}{\emph{1}}&  177.00&    -0.12&     4.95&     0.05&     2.53&     0.02&   84.87&    0.13& 1&      1&     20&      1\\
\textcolor{Gray}{\emph{2}}&  150.00&    -0.30&     4.60&     0.08&     2.62&     0.03&   97.83&    0.32& 1&     10&     37&     10\\
\textcolor{Gray}{\emph{3}}&  169.00&    -1.05&     5.54&     0.14&     2.96&     0.06&  112.10&    0.62& 1&      8&     34&      8\\
\textcolor{Gray}{\emph{4}}&  165.00&    -1.47&     8.12&     0.68&     6.23&     0.32&   66.10&    3.91& 1&      0&      0&      0\\
\textcolor{Gray}{\emph{5}}&  173.00&    -1.75&     0.00&     0.00&     0.00&     0.00&    0.00&    0.00& 0&      2&     21&      2\\
\textcolor{Gray}{\emph{6}}&  153.00&    -0.29&     8.19&     1.17&     0.00&     0.00&   54.30&    5.06& 0&      5&     27&      5\\
\textcolor{Gray}{\emph{7}}&  145.00&    -1.42&    10.50&     3.57&     0.00&     0.00&  101.42&   11.46& 0&      7&     33&      7\\
\textcolor{Gray}{\emph{8}}&  152.00&    -2.04&     6.69&     3.02&     0.00&     0.00&  138.33&   16.89& 0&      9&     36&      9\\
\textcolor{Gray}{\emph{9}}&  162.00&     1.63&     0.00&     0.00&     0.00&     0.00&    0.00&    0.00& 0&      4&     23&      4\\
\textcolor{Gray}{\emph{10}}&  191.00&     2.67&     0.00&     0.00&     0.00&     0.00&    0.00&    0.00& 0&      3&     22&      3\\
   \hline
\end{tabular}
\end{minipage}
\end{table*}

Table \ref{tab:components} lists the first ten components drawn from the eastern catalogue in order to illustrate the structure of the component table. The columns are defined as follows:\\
\noindent
(1) Identifier for the component formed from its HHMMSS.SS+/-DDMMSS.SS right ascension and declination position in J2000 coordinates.\\
\noindent
(2) Right ascension of the component in degrees.\\
\noindent
(3) Declination of the component in degrees.\\
\noindent
(4) The 1$\sigma$ uncertainty in right ascension in arcseconds.\\
\noindent
(5) The 1$\sigma$ uncertainty in declination in arcseconds.\\
\noindent
(6) Integrated flux density of the component in mJy.\\
\noindent
(7) The 1$\sigma$ uncertainty in the integrated flux density of the component in mJy.\\
\noindent
(8) Peak intensity of the component in mJy beam$^{-1}$.\\
\noindent
(9) The 1$\sigma$ uncertainty in the peak intensity of the component in mJy beam$^{-1}$.\\
\noindent
(10) Estimate of the local RMS noise at the position of the component in $\mu$Jy beam$^{-1}$.\\
\noindent  
(11) Spectral index ($\alpha$) of the component formed from an error-weighted mean of the pixel values in the spectral index mosaic over the extent of the component, or the extent of the survey restoring beam, whichever is larger.\\
\noindent  
(12) Deconvolved major-axis size of the Gaussian fitted to the component in arcseconds. A value of zero indicates that the source finder was unable to determine the intrinsic size, and the component is indistinguishable from being point-like.\\
\noindent
(13) The 1$\sigma$ uncertainty in the major axis of the component in arcseconds.\\
\noindent
(14) Deconvolved minor-axis size of the Gaussian fitted to the component in arcseconds. A value of zero indicates that the component is indistinguishable from being point-like.\\
\noindent
(15) The 1$\sigma$ uncertainty in the minor axis of the component in arcseconds.\\
\noindent
(16) Position angle measured east of north of the Gaussian fitted to the component in degrees.\\
\noindent
(17) The 1$\sigma$ uncertainty in the position angle of the component in degrees.\\
\noindent
(18) Boolean flag to indicate whether or not the component is likely to be reliably spatially resolved, i.e.~does the component lie above the upper envelope on Figure \ref{fig:resolved}.\\
\noindent
(19) Zero-indexed unique identifier for the component.\\
\noindent
(20) Zero-indexed unique identifier for the source.\\
\noindent
(21) Zero-indexed unique identifier for the island.\\

Note that there is not a direct one-to-one correspondence between a component having zeros in columns 12, 14 and 16, and a one in column 18. Both of these methods feature different statistical approaches, and the column 18 method in particular  is a somewhat blunt approach that is applied to capture the bulk properties of the source catalogue when determining source counts (see Section \ref{sec:bias} for details). When considering the morphological properties of individual sources, columns 12, 14 and 16 are likely to be more reliable, and the ratio of columns 6 and 8 is also a good indicator.

\section{Results and discussion}
\label{sec:discussion}

In this section the data products presented in Section \ref{sec:results} are examined. We verify the positional accuracy, the photometry and the in-band spectral measurements by making use of existing radio surveys and simulations. We present the differential source counts as derived from the survey, and present radio images of the four bright and extended sources in the survey area that are not included in the component catalogue.

\subsection{Comparison with existing 1.4 GHz radio observations of Stripe 82}
\label{sec:imgcomp}

\begin{figure*}
\begin{center}
\setlength{\unitlength}{1cm}
\includegraphics[width = \textwidth]{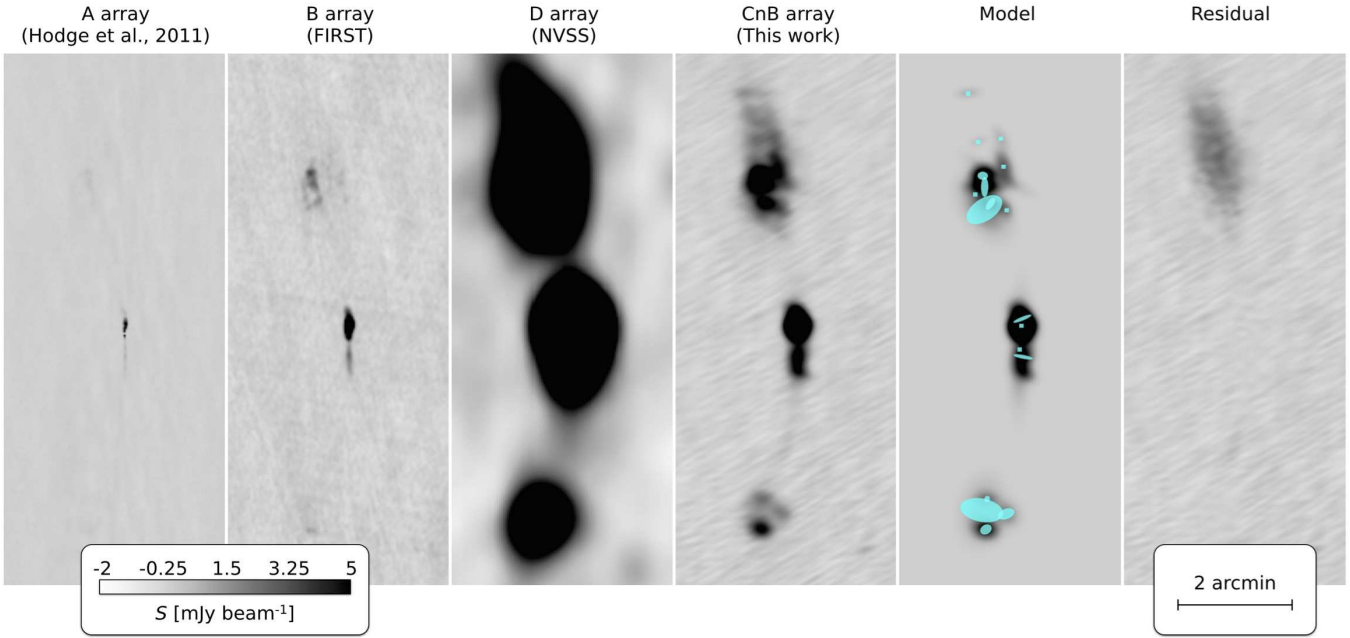}
\caption{A single radio galaxy is used to illustrate the continuum image products generated by this survey, and present a comparison to previous 1.4~GHz data of the Stripe 82 region. The source in question is a wide angle tail radio galaxy associated with SDSS J013412.84-010724.8 at a redshift of 0.078. The first three columns show the high resolution A-array data of \citet{Hodge2011}, the B-array FIRST data, and the D-array images from NVSS, with the CnB data presented in this paper in the fourth column. The greyscale is linear and runs from -2 to 5 mJy~beam$^{-1}$ for all panels as shown at the base of the figure. The fifth column shows a reconstruction of the source formed by convolving the catalogued point and Gaussian components (the cyan markers) with the restoring beam of the radio mosaics. Subtraction of the reconstructed model image from the data results in the residual image shown in the final column. Note the diffuse feature associated with the northern lobe that is not picked up by the source finder and is therefore absent from the catalogue.}
\label{fig:examplesource}  
\end{center}
\end{figure*}

The radio imaging of Stripe 82 presented in this paper is complementary with existing observations of the field at 1.4~GHz in terms of angular resolution and depth. Our observations lie between the A-array observations of \citet{Hodge2011} and the FIRST survey in terms of depth, and the CnB configuration means that they bridge the FIRST and NVSS datasets in terms of angular resolution and surface brightness sensitivity. The suitability of the new data for morphological classification of extragalactic radio sources is illustrated in Figure \ref{fig:examplesource} which presents the wide angle tail radio source associated with the galaxy SDSS J013412.84-010724.8 ($z$~=~0.078). The first three columns show the A-, B-, and D-array data with the CnB data in the fourth column. Much of the diffuse structure associated with the source is resolved out by the A- and B-array data, and while the NVSS image detects the large scale structure at high significance it is lacking the angular resolution required to disentangle core, jet and hotspot emission, all of which are readily identifiable in the CnB map.

The fifth column shows the morphology of the source as reconstructed by convolving the point (cyan squares) and Gaussian (cyan ellipses) catalogue components with the restoring beam used in the imaging, and the final column shows the residual of the model reconstruction subtracted from the Stokes-I data. The large diffuse feature associated with the northern lobe in the residual map is not detected and modelled by the source-finding software, and is thus absent from the components catalogue. The recovery of diffuse, low surface brightness emission is problematic for most source finders that typically make use of peak-finding algorithms. For this reason total source flux densities determined by summing catalogue components should be considered to be lower limits. The model and residual mosaics form secondary data products along with the total intensity and spectral mosaics. The residual image in particular should be the default product for any continuum stacking experiments \citep[e.g.][]{Zwart2014}.

\subsection{Extended sources that are excluded from the components catalogue}
\label{sec:sources}

\begin{figure*}
\begin{center}
\setlength{\unitlength}{1cm}
\includegraphics[width = \textwidth]{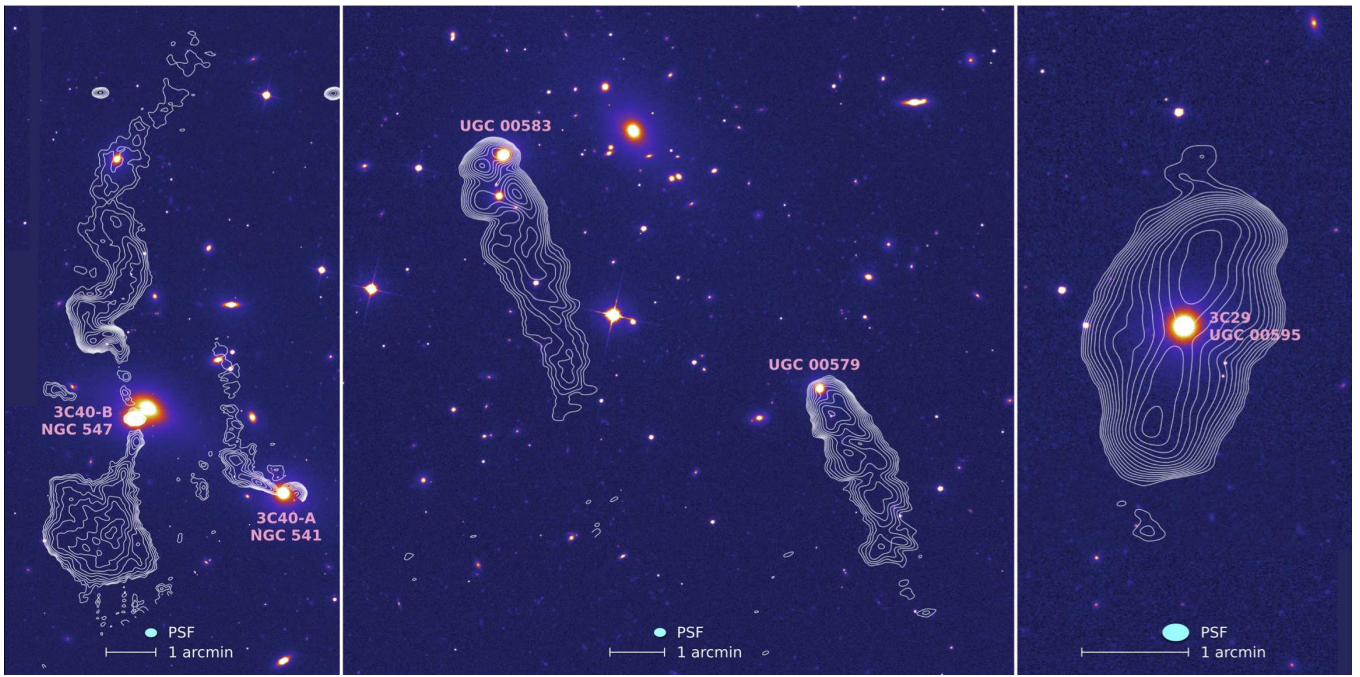}
\caption{Bright and extended radio galaxies that are not included in the component catalogue. These are 3C40-A and 3C40-B in the Abell 194 cluster (left panel), the twin narrow angle tail galaxies (centre panel) and the bright FR-I source 3C29 (right panel). The centre and right panels show sources that are all associated with the cluster Abell 119. The contours trace the total intensity radio emission with levels of (1, $\sqrt{2}$, 2, 2$\sqrt{2}$, 4, 4$\sqrt{2}$, 8, 8$\sqrt{2}$, ...)~$\times$~2 mJy beam$^{-1}$, overlaid on the SDSS $r$-band image. Arcminute scale bars and the size of the restoring beam are provided in each panel.}
\label{fig:extended}  
\end{center}
\end{figure*}

The survey area contains four bright and extended regions of emission for which {\tt PyBDSM} was unable to obtain a suitable point and Gaussian component fit in a reasonable amount of time. These sources are thus excluded from the component catalogue presented in Section \ref{sec:catalogue}. Total intensity contours for these sources are overlaid on the SDSS $r$-band image in Figure \ref{fig:extended}. The left panel shows the radio source 3C40, which is actually two distinct radio galaxies within the Abell 194 cluster \citep{Sakelliou2008}. 3C40-A is a narrow angle tail (NAT) source associated with NGC 541, and 3C40-B (the brighter and more extended of the two) is a wide angle tail (WAT) source associated with NGC 547. The remaining two panels of Figure \ref{fig:extended} show three sources that are associated with the cluster Abell 119. The centre panel shows two sources that are close to the cluster centre in projection. Higher resolution radio imaging \citep{Feretti1999} reveals these to be a pair of NAT sources. The final source in the right hand panel is 3C 29, a FR-I \citep{Fanaroff1974} source in the periphery of Abell 119.

\subsection{Survey sensitivity: thermal noise vs effective noise}
\label{sec:sensitivity}

Figure \ref{fig:rms} shows the measured spatial variation in the image sensitivity expressed as the RMS image noise, the determination of which is described in Section \ref{sec:catalogue}. As usual with a primary beam corrected radio image the effective noise is raised at the edges of the field as the off axis source flux densities attenuated by the primary beam response are corrected to their intrinsic values and the image noise is correspondingly raised. Aside from this, regions of elevated noise are otherwise coincident with bright or extended sources where incomplete deconvolution (primarily due to calibration deficiencies) results in residual PSF sidelobes that are interpreted as an effectively higher noise floor.  

\begin{figure}
\centering
\includegraphics[width=\columnwidth]{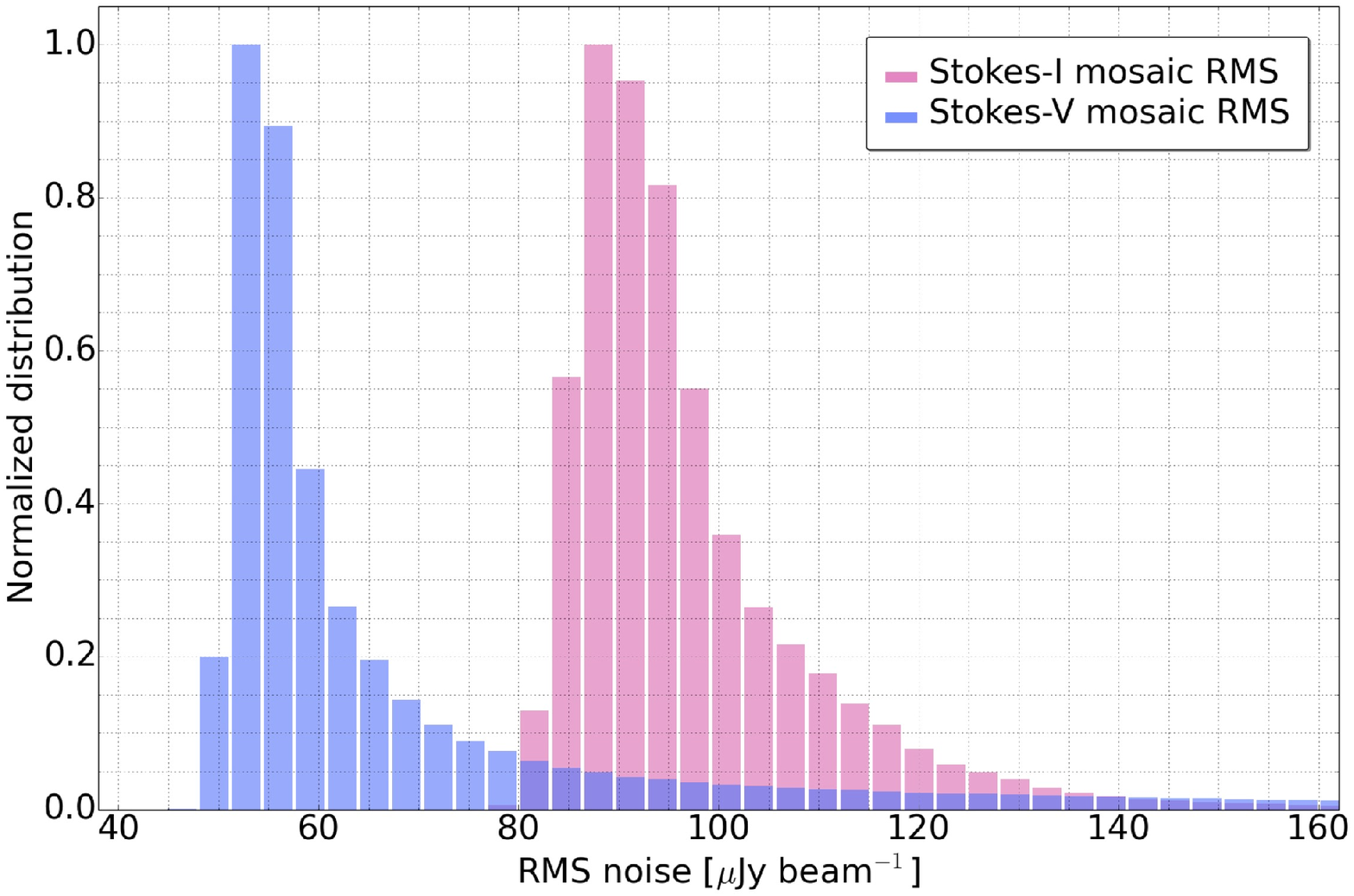}
\caption{Normalized histograms of the RMS values from Figure \ref{fig:rms} as determined by the {\tt PyBDSM} source finder and those measured via the formation of Stokes-V images of the individual survey pointings. The Stokes-V measurements peak at 52 $\mu$Jy beam$^{-1}$, however the effective survey ``noise'' as estimated by the source finder is higher, with the distribution peaking at 88~$\mu$Jy beam$^{-1}$. See text in Sections \ref{sec:selfcal} and \ref{sec:sensitivity} for further details.}
\label{fig:cumu_rms}
\end{figure}

Figure \ref{fig:cumu_rms} shows a normalized histogram of the pixel RMS values shown in Figure \ref{fig:rms} (pink), as well as those measured from an equivalent mosaic formed from the Stokes-V images mentioned in Section \ref{sec:selfcal} (blue). The RMS values of the Stokes-V mosaic are assumed to provide a reasonable measurement of the thermal noise of the observations. The Stokes-V images are largely empty: any residual peaks in emission are coincident with bright sources, occurring due to (uncorrected) instrumental leakages and manifesting itself as a small fraction of the Stokes-I flux density. The distribution of the Stokes-V measurements peaks at 52~$\mu$Jy~beam$^{-1}$, very close to the target thermal noise of 50~$\mu$Jy~beam$^{-1}$ for which the survey was designed. The values measured in the Stokes-I mosaic are higher, with the distribution peaking at 88~$\mu$Jy~beam$^{-1}$. The difference between the thermal and effective noise is  
likely due to the following effects: (i) the conservative and automated approach to deconvolution (see Section \ref{sec:selfcal}) resulting in residual sidelobe structures, particularly around extended sources; (ii) calibration deficiencies for pointings close to very strong sources; (iii) the significant number of pixels that form the perimeter of the survey area having elevated noise levels due to the primary-beam correction.
 
\subsection{Astrometric accuracy}
\label{sec:astrometry}

The uncertainties in the measured position of a radio source consist of a statistical component related to the signal to noise ratio (SNR) of the detection as well as the angular resolution of the instrument, and a component associated with systematic astrometric frame errors present in the observation. The latter is generally due to calibration deficiencies \citep{condon1997}. Systematic offsets in the positional measurements of the sources in the components catalogue, as well as an estimate of the level of statistical uncertainty, can be determined by cross matching the source positions with those determined in independent radio surveys. Systematic offsets are typically investigated by selecting sources that are bright enough such that the statistical component is negligible (e.g.~typical phase calibrators) and comparing their measured positions against those obtained with Very Long Baseline Interferometry (VLBI) measurements. 

To verify the astrometry of the survey we compare the catalogue positions to matched sources from the A-array data of \citet{Hodge2011} and the B-array data from the FIRST survey \citep{Becker1995}, both of which have superior angular resolution to our CnB data. The component catalogue described in Section \ref{sec:catalogue} is filtered to include only entries where the island identifier is associated with a single component. This produces a subset of components that does not contain very extended or morphologically complex emission (with the exception of the four sources shown in Figure \ref{fig:extended} and discussed in Section \ref{sec:sources}). Component lists from each of the external surveys are then cross referenced with the filtered CnB sample in order to find the nearest associated component, although the cross-match is rejected if the nearest component is separated by more than 5 arcseconds to exclude spurious detections and components that are not detected in both surveys. This results in 2,626 and 2,773 cross matches for the external A- and B-array surveys respectively.

\begin{figure}
\centering
\includegraphics[width=\columnwidth]{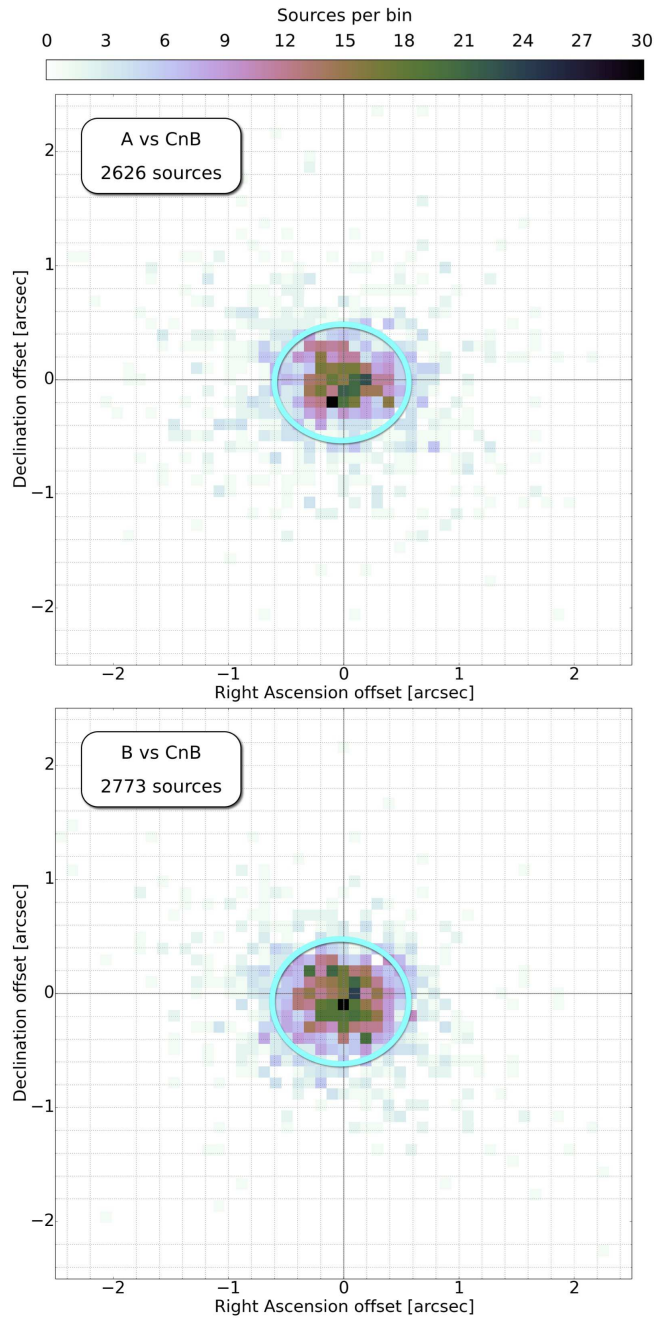}
\caption{Measured positional differences between catalogued sources from the survey presented in this paper, and those detected by Hodge et al.~(2011; upper panel) and the FIRST survey (Becker, White \& Helfand., 1995; lower panel). The points on the above figure relate only to those entries in the component catalogue that are isolated, i.e.~entries where the island identifier belongs only to one component. The images show the two dimensional histograms (51~$\times$~51 bins) of the offsets in right ascension and declination. The ellipses are centred on the mean offset positions and their extents show the corresponding 1$\sigma$ scatter in the points in right ascension and declination between the CnB survey and the corresponding external survey. The extents of the panels above are one half of the minor axis of the restoring beam used in the CnB survey.}
\label{fig:astrometry}
\end{figure}

The offsets in right ascension and declination are determined for matched components, and the distribution of these offsets is shown as a two-dimensional histogram with 51~$\times$~51 bins in Figure \ref{fig:astrometry}, with the A- and B-array comparisons in the upper and lower panels respectively. The mean ($\pm$1 standard deviation) values of the distributions in (RA, Dec) in units of arcseconds are ($-$0.018~$\pm$~1.171, $-$0.023~$\pm$~1.018) and ($-$0.03~$\pm$~1.186, $-$0.069~$\pm$~1.094) for the A and B comparisons respectively. The cyan ellipses on each panel are centred on the mean offset position, and the extent of the major and minor axes are one standard deviation of the distribution of offsets in RA and Dec. The distributions also confirm that the $<$5 arcsec separation criterion is not biasing the results.

For the A-array and B-array comparisons respectively, the 1$\sigma$ scatter values are 0.8 arcsec and 0.89 arcsec in right ascension and 0.5 arcsec and 0.6 arcsec in declination. The primary source of the increased scatter in the CnB measurements is likely to be the angular resolution of the survey, noting the increased right ascension offset is consistent with the PSF being elongated in that direction. Scaling the RMS offsets reported by \citet{Hodge2011} by the ratios of the angular resolutions of the A and CnB surveys results in higher scatter than is reported here, however \citet{Hodge2011} reason that bandwidth smearing contributes significantly to their uncertainties. This does not significantly affect our measurements due to our choice of imaging method. Measuring the mean value of the offsets as a function of flux density cut off reveals no systematic bias. Note that the positional uncertainties in the component catalogue are derived from the uncertainties in the fit performed by {\tt PyBDSM}, and are therefore purely statistical.

\subsection{Photometric accuracy}
\label{sec:photometry}

The calibration of the absolute flux density scale against the model of a strong calibrator source provides the VLA with an excellent measurement accuracy of order one percent \citep{Perley2013}. There are however factors external to the absolute calibration that can bias the flux scale accuracy, e.g.~deficiencies with the subsequent referenced and self-calibration, deconvolution effects and RFI.

\begin{figure}
\centering
\includegraphics[width= 0.87 \columnwidth]{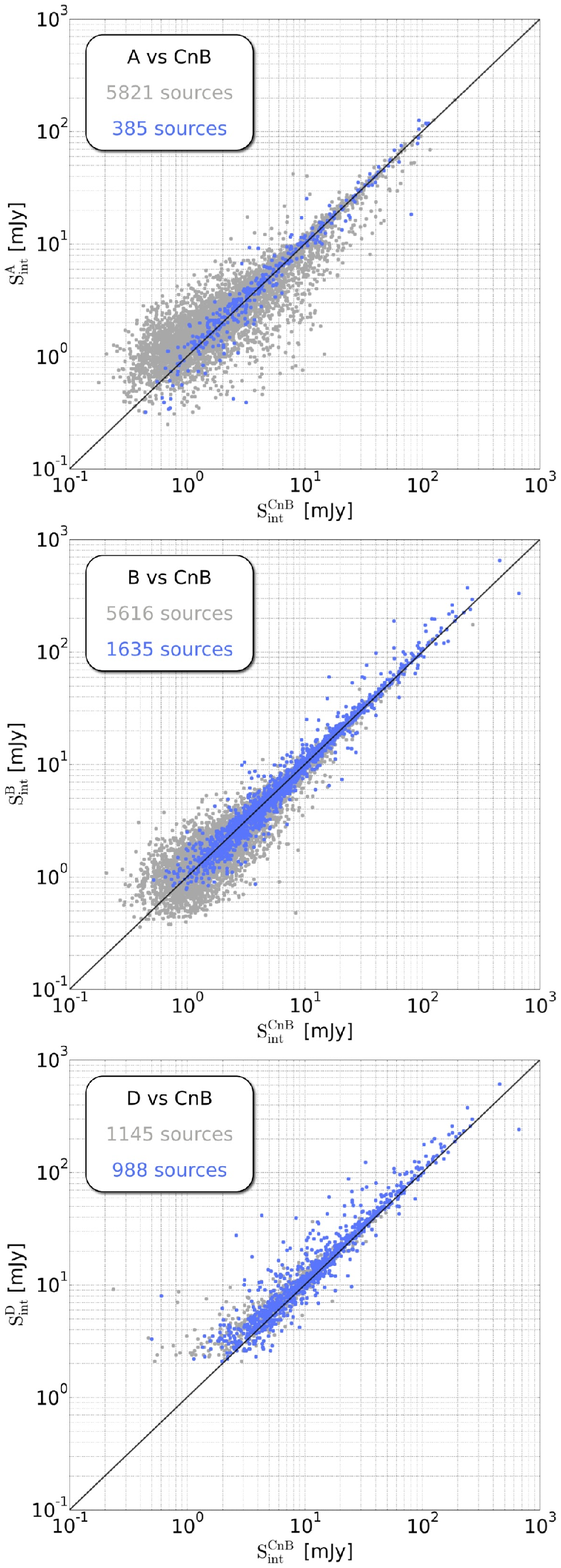}
\caption{Comparison of the integrated flux density measurements of our survey with those of existing surveys over the same region, namely the A-array observations of \citet{Hodge2011}, the B-array observations of the FIRST survey \citep{Becker1995} and the D-array NVSS data \citep{Condon1998}. As with Figure \ref{fig:astrometry} only single component islands are selected for comparison in order to minimise the effects of the differing sensitivities to extended structure. The number of matched components is shown on each panel, with the total number shown in grey and the unresolved (in both the CnB catalogue and the reference catalogue) matches shown in blue.}
\label{fig:photometry}
\end{figure}

Our first check is to compare the flux density measurements of the catalogued components to those of existing surveys. Again we employ the A-array data of \citet{Hodge2011} and the FIRST survey B-array data of \citet{Becker1995}, and additionally the D-array data from the NVSS \citep{Condon1998}. As with the positional matching, islands that contain only a single component are selected from the catalogue, and matches are rejected if the separation between the CnB position and that of the comparison survey exceeds 5 arcseconds. Applying these criteria results in 5,821, 5,616 and 1,145 matches from the A-, B- and D-array data, respectively. These subsets are plotted as grey points on Figure \ref{fig:photometry}, which shows the comparison integrated flux density measurements against those of the CnB components. The 1:1 ratio line is indicated on the plot, and there is increased scatter about this line with decreasing flux density. This is to be expected as the image noise becomes a more significant fraction of the source flux density at fainter levels. A 5$\sigma$ source that is observed multiple times with identical observations will appear to be 20 percent variable by definition.

The three external surveys being used for this comparison have markedly different angular resolutions and sensitivities to extended structure. Despite selecting only isolated components, the differences that this introduces to the flux density comparisons are also evident in Figure \ref{fig:photometry}. For example, the points tend towards a flux excess at the faint end in the high angular resolution A-array data in the upper panel. A similar effect is noted by \citet{Hodge2011} which is interpreted in terms of a combination of resolution effects and Malmquist bias. A component that is resolved in one survey will exhibit a lower peak flux density than its corresponding unresolved measurement. Such a source will be picked up by a peak-finding source detection algorithm only if a positive noise contribution pushes the peak over the threshold. Such sources are therefore biased to high flux densities in the higher resolution data. The turn-up in the distribution of grey points due to this effect is lessened in the B-array data, as expected.

The effects of the mismatched angular resolutions can be further illustrated by requiring components to have peak to integrated flux density ratios of between 0.9 and 1.1 in both samples. This selects sources that are unresolved (or barely resolved) by both the CnB survey and the external reference survey. These components are plotted as blue points on Figure \ref{fig:photometry}, and they sit much closer to the 1:1 line. There is a hint of an excess flux density bias in the faint end of the NVSS measurements. This could also be a resolution effect even for sources that are selected to be point like as extended emission seen by the D-configuration of NVSS may be being resolved out in the CnB data (cf. Figure \ref{fig:examplesource}).

An additional source of photometric error occurs due to CLEAN (or snapshot) bias. This is an incompletely understood effect related to the deconvolution process, typically manifesting itself as a systematic underestimation of the source brightnesses \citep{Condon1998,Becker1995}. The deficit is typically found to be fixed down to some flux density threshold, below which it becomes flux dependent and has even been shown to affect sources below the thermal noise limit of the images \citep{White2007}. The magnitude of the effect appears to be coupled to the PSF sidelobe level, with the strong linear features in the VLA snapshot PSF thought to be particularly problematic, and unconstrained deconvolution where sidelobe features are more likely to enter the clean model can also cause increased suppression of the field sources. Few attempts have been made to investigate the problem in the context of high fractional bandwidth continuum observations although \citet{Helfand2015} note that the bias appears to correlate with RFI levels.

The deconvolution strategy described in Section \ref{sec:selfcal} was designed to be conservative in order to minimise the effects of clean bias. We estimate the level at which it is present in the final images via the standard technique of injecting artificial sources of known flux density into the calibrated data, re-imaging and comparing the recovered flux density to the known intrinsic flux density. The artificial sources are flat-spectrum point sources placed at random positions within the band centre half power point of the primary beam, with flux densities that span twelve logarithmically-spaced values between 0.32 and 158 mJy inclusive. Such sources were injected into the Measurement Sets of 500 pointings which were then re-imaged, a total of 6,000 independent simulations. If the flux density of the test source exceeded the 1~mJy catalogue limit of the FIRST survey then the cleaning mask was also modified to include that source. These simulations were performed on the Galaxy\footnote{{\tt http://www.pawsey.org.au/our-systems/\\

galaxy-technical-specifications/}} system at the Pawsey Supercomputing Centre.

\begin{figure}
\centering
\includegraphics[width=\columnwidth]{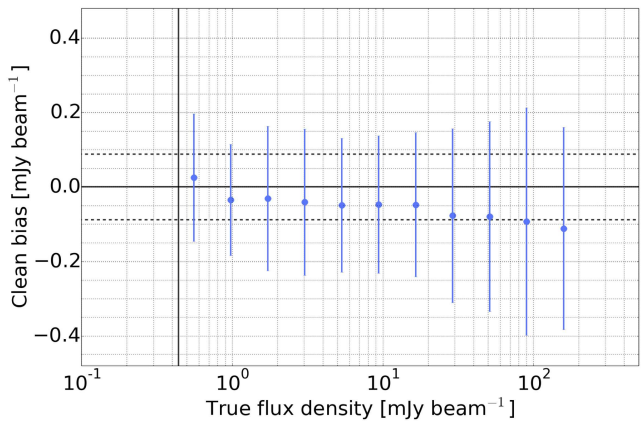}
\caption{The results of the simulations used to quantify the level of clean (or snapshot) bias in the survey images. The x-axis shows the known true flux density of the flat-spectrum point source that was injected into the data. The y-axis shows the measured clean blas, i.e.~the measured minus the true flux density. The dashed horizontal lines show $\pm$1$\sigma$ where $\sigma$ is the typical effective RMS noise in the survey (88~$\mu$Jy~beam$^{-1}$) and the vertical line shows the typical detection threshold at five times this value. Statistics on the clean bias are gathered by running the simulation on 500 of the single pointing Measurement Sets from the survey data for each flux density value. The points show the mean value of the clean bias recovered from the simulations, and the error bars show $\pm$1 standard deviation of the distribution.
}
\label{fig:clean_bias}
\end{figure}

Figure \ref{fig:clean_bias} shows the results of these simulations, with the true flux density of the artificial flat-spectrum source inserted into the data along the x-axis, and the clean bias (i.e.~the recovered minus the intrinsic flux density) on the y-axis. The points show the mean values from each flux density bin as measured from the 500 different Measurement Sets into which the artificial sources were inserted. The error bars show $\pm$1 standard deviation of the values in each flux density bin. The dashed horizontal lines are $\pm$1$\sigma$ where $\sigma$ is the typical RMS noise in the survey (88~$\mu$Jy~beam$^{-1}$). The vertical line shows the typical detection threshold at five times the typical RMS noise. As can be seen the lowest bin is below the typical source detection threshold. Since the position of the source was known, a forced fit to the component at that location could be made.

The clean bias expressed as a percentage of the component flux density in the second lowest bin is 5 percent. This is therefore the \emph{average} worst-case error for the clean bias in the \emph{catalogued} component flux densities. The fractional clean-bias errors are $<$1 percent by the fifth flux density bin. Persons making use of the survey images in the sub-5$\sigma$ regime should be mindful that the clean bias errors may lie at the few tens of percent level. The clean bias manifesting itself as an excess at the faint end is also not typically seen in such investigations. This may be an artefact of the MT-MFS algorithm used to deconvolve the broadband data, an effect not seen in previous narrowband studies. Note that although the bias appears to rise significantly with increasing flux density, when expressed as a fraction of the component flux it becomes increasingly insignificant, dropping from 3 percent in the third bin to 0.07 percent in the highest bin. We make no attempt to account for the increase in the fractional-clean bias error that will result from the primary-beam correction: the actual catalogued flux-density measurement for a component will be a weighted sum of its measurements in all the pointings that contribute to the linear mosaic at that position. As this sum is weighted by the assumed noise variance of the image (i.e.~it is coupled to the assumed primary beam pattern) the component flux density measurement will be dominated by those images where the component lies closest to the pointing centre. However in addition to this, and as we will shortly discuss, the clean bias effect appears to vary significantly between pointings, so there is no way to rigorously quantify the bias in terms of the radial separation of a component from the contributing pointing centres.

Unmasked deconvolution with fixed numbers of iterations on a limited subset of the data, followed by a smaller scale round of clean bias simulations, demonstrated a clear advantage for our mask-based deconvolution approach in terms of faithful component photometry. Nonetheless, a significant clean bias effect remains despite our conservative approach. There appears to be no obvious way to mitigate this, short of fully interactive and iterative deconvolution of the data, complete with simulations that monitor the clean bias, allowing an optimum deconvolution strategy to be implemented on a per-pointing basis. The usual statistical approach to quantifying it and correcting it via the injection of artificial sources, as has been implemented here, clearly falls short: the error bars indicate large amounts (typically $\pm$0.2 mJy beam$^{-1}$) of scatter in the measured clean bias. The apparent dependence of the bias on RFI (Helfand, White \& Becker, 2015) has already been mentioned, and residual low level RFI that has been missed by the automatic and manual flagging operations may be present and contributing to the spread. The significant spread in the extents of the main lobes of the per-pointing PSFs (Section \ref{sec:selfcal}) gives another hint, as it is suspected that clean bias is coupled to the PSF sidelobe levels. Despite the very narrow declination range of our survey area, dynamic scheduling of the observations demands a fairly generous range of acceptable hour angle (or local sidereal time) ranges in order to maximise the chances of a Scheduling Block being selected from the queue. The end result is that the elevation of the field can change significantly between observations, and this gives rise to significantly different PSF sidelobe structures. Optimally rather than dynamically scheduled observations are clearly advantageous for large area radio surveys where uniformity of the data products is highly desirable.

A deeper investigation of the clean bias effect in broadband snapshot VLA surveys will be pursued using these data, examining for correlations between the bias and peak sidelobe levels, amount of RFI, and by searching for a frequency dependence by conducting sub-band rather than full-band imaging, with corresponding sets of artificial source simulations. The use of the MT-MFS algorithm, which models the spectral behaviour of clean components as a polynomial pegged to a reference frequency at the band centre, almost certainly plays a role, and the biases that this introduces in the frequency domain is examined in the next section.

\subsection{Spectral index and curvature}
\label{sec:alphas}

\begin{figure}
\centering
\includegraphics[width=\columnwidth]{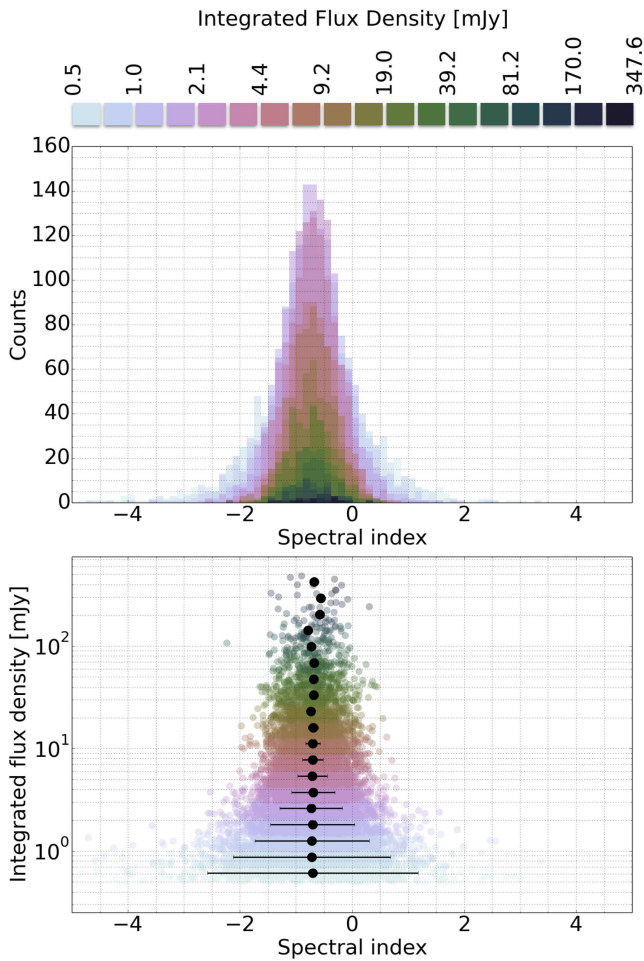}
\caption{Spectral index distribution made to illustrate the effect of image noise on the recovered $\alpha$ values, and simulate the distribution of a two-point measurement made by splitting our data into two 512~MHz sub-bands. Catalogued total intensity flux densities at 1.5~GHz are turned into measurements at 1.25 and 1.75~GHz by assigning a spectral index drawn from a normal distribution with a mean of $-$0.71 and a standard deviation of 0.38 \citep{Mauch2013}. This pair of flux density measurements is then perturbed by an appropriate noise value based on the true local RMS of the component. The values of $\alpha$ measured from the noisy flux density measurements are plotted above. Median values of $\alpha$ are plotted as black points, the error bars for which are the median values of the errors in $\alpha$ for that flux density bin. The main purpose of this figure is to illustrate the significant scatter in $\alpha$ that is introduced purely as a result of the SNR of the component decreasing.}
\label{fig:two_point_noisy}
\end{figure}

Estimates of the spectral index ($\alpha$) and curvature ($\beta$) are made for each component in the catalogue by extracting pixels from the $\alpha$ and $\beta$ mosaics at the position of the component, and over an area that is either the fitted size of the Gaussian component or the area of the restoring beam used to form the mosaic, whichever is larger. A corresponding area is extracted from the spectral index error mosaic and $\alpha$ and $\beta$ values are assigned to each component based on the error-weighted mean of the pixels in the extracted area. Note that the spectral index error maps produced by CASA are not necessarily formal uncertainties in the value of $\alpha$, and are intended to be used (as they are here) as a coarse estimate of the reliability of the value of $\alpha$ at any given position in the map. The error estimates are derived from the statistics of the residuals in the Taylor term images, and these are propagated into the calculation of $\alpha$ via Equation \ref{eq:alpha}, resulting in a map of the spectral index error.

The usual approach to determining the spectral index of a radio source is to observe it using two or more separate observations at different radio frequencies, preferably matched in resolution and depth, and fit a spectral index to the flux density measurements \citep[e.g.][]{Mauch2003,Prandoni2006,Ibar2009,Randall2012,Mauch2013,Huynh2015}. Noise in the independent flux density measurements propagates into the spectral index measurement, and reliable spectral index determination requires both high signal-to-noise ratio flux-density measurements, as well as observations that are maximally separated in frequency but remain within the regime where the spectrum is adequately described as a power law (this separation in frequency is sometimes referred to as the spectral 'lever arm'). Catalogues of large numbers of objects derived from radio surveys are typically extracted down to flux density limits of 5$\sigma$, and matching two pairs of such catalogues introduces significant scatter in the spectral index distribution from noise considerations alone. These spectral index distributions are thus best suited to studying the \emph{average} spectral index behaviours of large numbers of radio sources, for example median spectral index as a function of flux density, population type, etc.

The notion that the new generation of broad-band radio interferometers can make accurate in-band spectral index measurements is an attractive one in terms of making economical use of telescope time. Two approaches are available here, namely (i) imaging the data in matched resolution sub-bands and deriving spectral index measurements in a method analogous to the dual-observation approach described in the paragraph above \citep[e.g.][]{heywood2016}; and (ii) using broadband deconvolution algorithms to image the entire band, fitting an estimate of the spectral behaviour of each component in the process. The latter method has the distinct advantage of providing a single image that has the depth afforded by the broad bandwidth, the angular resolution of the approximate band centre frequency and the PSF sidelobe benefits provided by MFS gridding. It is this method that we adopt and verify here, using the MT-MFS algorithm as described in Section \ref{sec:selfcal}. 

By way of a benchmark, Figure \ref{fig:two_point_noisy} presents a simulation of the spectral index distribution that would be obtained by dividing the 1--2 GHz of bandwidth into two parts and determining $\alpha$ from flux density measurements at 1.25 and 1.75 GHz. This is achieved by taking the catalogued component flux densities at 1.5~GHz together with their local RMS measurements. A spectral index is assigned to each component drawn from a normal distribution with a mean of $-$0.71 and a standard deviation of 0.38, adopting the distribution measured by \citet{Mauch2013}. The assigned spectral index value is used to compute 1.25 and 1.75~GHz flux densities based on the catalogued 1.5~GHz value, and these are then perturbed by two independent noise values drawn from a normal distribution centred on zero with a standard deviation of $\sqrt{2}$ times the local RMS value. This simulates the increase in noise that would result from imaging the band in two 512~MHz chunks, assuming equal sensitivity across the VLA's 1--2 GHz band. An $\alpha$ value is computed from the two noisy measurements and the resulting distribution is shown in Figure \ref{fig:two_point_noisy}. The upper panel shows a histogram of the $\alpha$ values, colour coded by where the component flux density lies in a set of logarithmically-spaced integrated flux density bins, as indicated at the top of the figure. The lower panel shows the distribution of the component integrated flux densities as a function of $\alpha$, with the black points showing the median $\alpha$ measurements in each flux density bin, with an error bar corresponding to the median error in $\alpha$ for the values in that bin. An error value for $\alpha$ for each individual point is computed by propagating the simulated flux density errors in the measurement at each frequency. Note that as the flux density decreases so does the SNR, and this introduces scatter into the $\alpha$ measurements that goes way beyond that of the assumed intrinsic $\alpha$ distribution. The SNR-related errors start to becoe significant below about 1 mJy.

\begin{figure*}
\begin{center}
\setlength{\unitlength}{1cm}
\includegraphics[width = \textwidth]{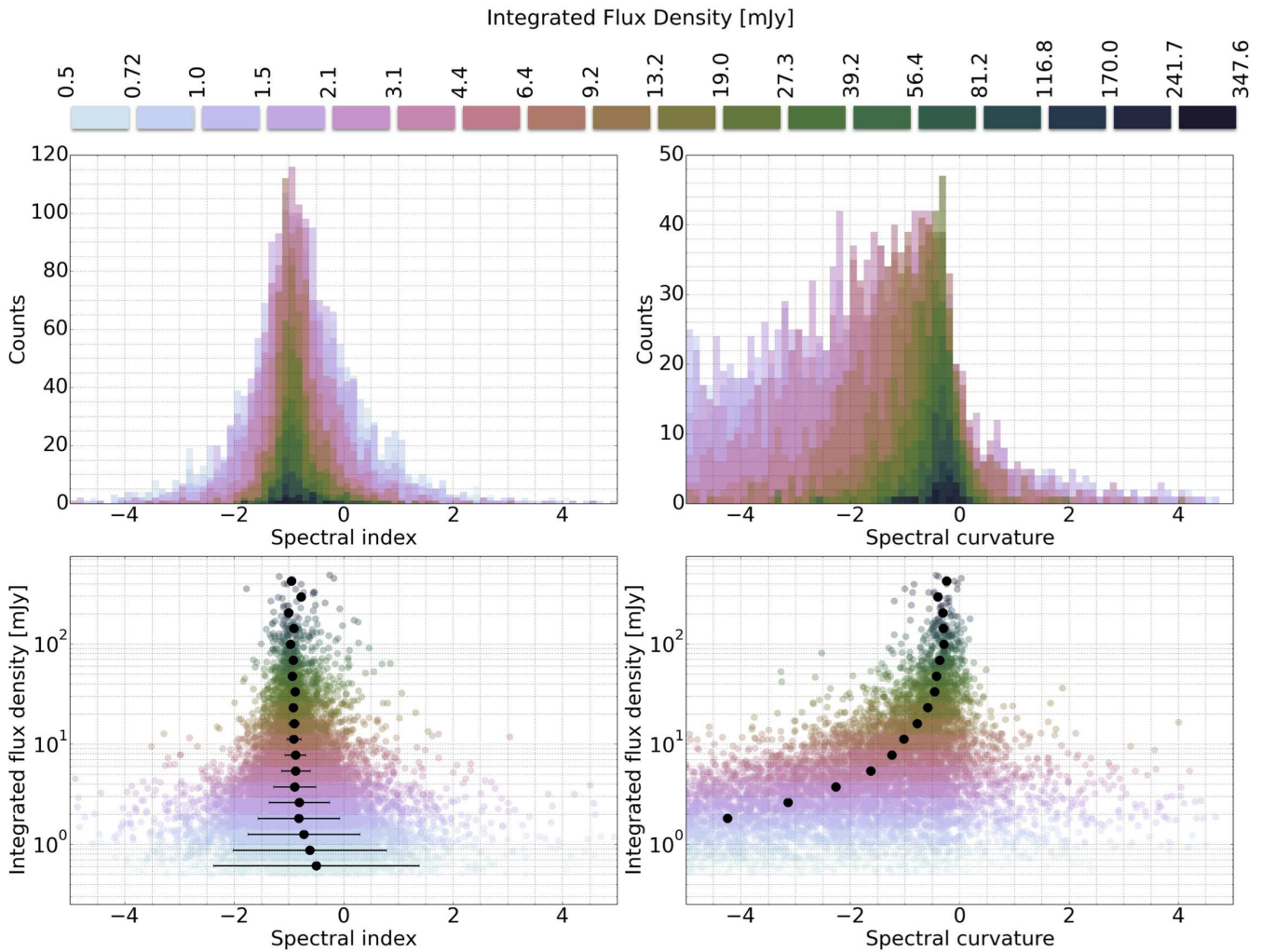}
\caption{The upper row shows histograms of the in-band $\alpha$ (left column) and $\beta$ (right column) values for each component, as derived from the third-order MT-MFS imaging. The lower row shows plots of the integrated flux density of each component as a function of $\alpha$ (left column) and $\beta$ (right column). All points are coloured according to where the component integrated flux density lies in a set of logarithmically-spaced bins as indicated at the top of the figure. The black points in the lower panels indicate the median $\alpha$ and $\beta$ measurements for each integrated flux density bin. Error bars on the median $\alpha$ values are those derived for Figure \ref{fig:two_point_noisy}, intended to provide an indication of the broadening of the distribution due to SNR considerations. The $\alpha$ values indicate a trend towards flatter spectra with decreasing flux density, whereas the $\beta$ values rapidly tend towards unreasonable spectral curvatures as the component flux densities decrease.}
\label{fig:alphas}  
\end{center}
\end{figure*}

The distribution of the spectral indices ($\alpha$) in the catalogue (measurement of which is described in Section \ref{sec:catalogue}) are presented in Figure \ref{fig:alphas}, along with the spectral-curvature measurements ($\beta$) which were extracted using the same method. As per Figure \ref{fig:two_point_noisy}, the upper panels show histograms of the $\alpha$ and $\beta$ values, colour coded by the component flux density, as indicated at the top of the figure. Broadening of the distribution with decreasing flux density is again evident, likely due to SNR considerations. The lower panels show the distribution of the component integrated flux densities as a function of $\alpha$ and $\beta$, with the black points showing the median $\alpha$ and $\beta$ measurements in each flux density bin. Error bars on the median values of $\alpha$ are those derived from the simulation used to produce Figure \ref{fig:two_point_noisy}. These show the approximate error introduced purely as a function of SNR, and independent of the underlying true spectral index distribution. The scatter introduced by decreasing SNR starts to become significant below about 1 mJy.

Our first check is to measure the spectral index of components matched with the quasar positions from the Stripe 82 DR7 quasar catalogue \citep{Shen2011}. This gives a median spectral index of $-$0.4 for compact (S$_{peak}$/S$_{int}$ $>$ 0.95) radio components compared to $-$0.58 for the components with extended emission. A Kolmogorov-Smirnov test rejects the hypothesis that the two spectral index samples are drawn from the same distribution ($p$-value $<$2 percent).

There is an apparent flattening of the median spectral index with decreasing flux density, as quantified in Table \ref{tab:medalphas} and shown in Figure \ref{fig:alphas}. The spectral behaviour of radio sources as a function of their flux density has been investigated by several authors. Studies at the bright end are best done with shallow, large-area surveys \citep[e.g.][]{Massardi2011} where a shift between flat spectrum quasars and radio galaxies with steeper spectra is seen at flux densities above $\sim$100~mJy. Moving to fainter flux density levels, \citet{Mauch2003} compare the 843 MHz Molonglo Observatory Synthesis Telescope (MOST) observations of the Sydney University Molonglo Sky Survey (SUMSS) to the 1.4 GHz NVSS measurements. A shift towards flatter spectra with decreasing flux density is observed, with a median spectral index of $-$0.89 above 50 mJy, increasing to $-$0.77 below 20 mJy, likely due to the shift from steep-spectrum Fanaroff-Riley Type-I radio galaxies \citep{Fanaroff1974} to lower luminosity flat spectrum AGN. For comparison, the median spectral index in our catalogue above 50 mJy is $-$0.92, and below 20 mJy it is $-$0.79, consistent with existing dual-frequency measurements.

Figure \ref{fig:alphas} shows a gradual shift towards a flatter median spectral index with decreasing flux density, and the presence of such a trend is not consistently reported in the literature. Previous studies of the spectral indices of sources at $\sim$mJy levels typically involve deeper surveys covering a few primary beam areas, and typically compare 1.4 GHz observations made with the VLA or ATCA with either lower frequency 325, 610 MHz GMRT or 843 MHz MOST observations, or higher frequency 5~GHz VLA or ATCA data. Some studies measure a flattening, e.g.~\citet{Prandoni2006} compare 1.4 and 5 GHz observations, and observe a median spectral index shift from $-$0.71 above a 1.4 GHz flux of 4~mJy to $-$0.53 for the components below this threshold. Authors who claim no evidence for spectral flattening include \citet{Ibar2009}, who present a spectral index measurement of between $-$0.7 and $-$0.6 for a source sample with 1.4 GHz VLA fluxes between $\sim$100~$\mu$Jy and 10 mJy, derived from well-matched GMRT observations at 610~MHz. For comparison, our median spectral index in this flux range is $-$0.8. \citet{Randall2012} find ``no statistically significant evidence for'' (but do not rule out) a systematic flattening based on 843 MHz MOST and 1.4 GHz ATCA observations. \citet{Huynh2015} present a median spectral index of $-$0.58 by cross matching a sample of 187 point sources observed with ATCA at 5 GHz with high resolution 1.4 GHz VLA data \citep{Miller2013}. These median measurements are flatter then the ones we present in Figure \ref{fig:alphas}, however this is likely a result of comparing a 1.4 GHz survey with a second survey at a higher frequency. At frequencies $\sim$10 GHz rest-frame and above, and faint flux densities, one would expect a typical galaxy to have a flatter radio spectrum as thermal (free-free) processes begin to contribute to the radio output at an increasing fraction compared to non-thermal synchrotron emission \citep{Whittam2013,Murphy2015}.

The spectral curvature of extragalactic radio sources is significantly less well studied than the typical spectral index, particularly for faint sources. While certain rare objects are known to exhibit extremely curved, peaked spectra \citep[e.g.][]{Callingham2015}, the amount of curvature exhibited by a typical synchrotron spectrum between 1--2 GHz is expected to be very slight. Guglielmino (2013) examines the spectra of 79 radio sources in the Lockman Hole between 345 MHz and 1.4 GHz. In that study, curvature is defined as the difference between the spectral index measured between 345 and 610 MHz, and that measured between 610 MHz and 1.4 GHz. The mean and median spectral curvatures for the sample are both consistent with zero. (We note however that the median $\alpha$ value shifts from $-$0.82 in the sample with 150 MHz fluxes above 60 mJy to $-$0.65 in the sample below 60 mJy.) In any case, MT-MFS is only expected to be able to fit meaningful $\beta$ values for typical (i.e.~only slightly curved) radio source spectra at very high SNRs ($\gtsim$100). The increasingly significant curvature with decreasing flux density present in our catalogue is suspicious, and will be discussed further below.

A truly consistent picture is yet to emerge from these many observations, however there are clearly many selection biases at play, such as: shot noise effects in the smallest-area, deepest observations capturing a mixture of source populations; the ability of higher-resolution imaging to separate core-jet sources into individual components that are blended in lower resolution studies; and differing survey depths at the two frequencies biasing the matched samples towards steep-spectrum sources at the faint end. The in-band spectral index measurements that are possible with new and upgraded broadband radio telescopes may offer considerable advantages in terms of minimising the differences present when determining spectral index measurements from observations made at two widely separated frequencies with two different instruments, for example resolution mismatches and different calibration systematics. In this paper we have opted for use of the MT-MFS deconvolution method to estimate the spectral properties of the components. However, the reliability of these measurements must be determined.

\begin{figure*}
\begin{center}
\setlength{\unitlength}{1cm}
\includegraphics[width = \textwidth]{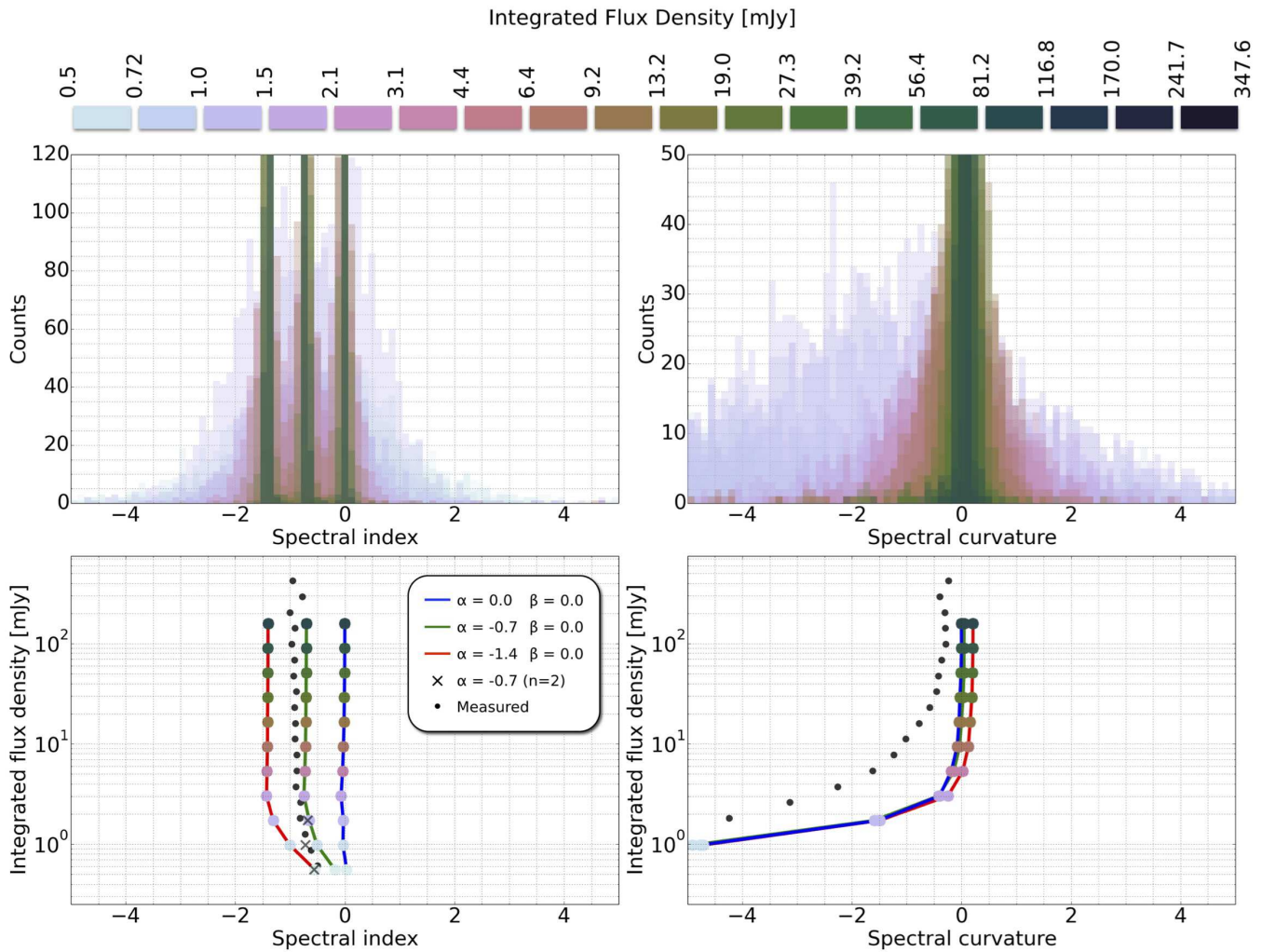}
\caption{The upper row shows histograms of the recovered values of $\alpha$ and $\beta$ derived from approximately 25,000 imaging simulations where sources of known spectral shape were inserted into the VLA data. The sources had $\alpha$ values of either $-$0.7, $-$1.0 or $-$1.4, and no spectral curvature ($\beta$~=~0) in each case. These three $\alpha$ values are shown by the coloured lines (as per the inset legend) on the lower plots which show the behaviour of $\alpha$ and $\beta$ as a function of the flux density of the artificial source. The black points show the median $\alpha$ and $\beta$ values as derived from the observations and presented in Figure \ref{fig:alphas}, and in the case of the spectral indices, Table \ref{tab:medalphas}. This figure shows that the observed trend towards flatter spectra with decreasing flux density is at least partially an artefact of the MT-MFS algorithm as it was applied to these data, and that the curvature values should be disregarded. The crosses in the lower left panel show a smaller scale simulation whereby $\alpha$~=~$-$0.7 sources in the lowest three flux density bins were imaged using only a second-order Taylor expansion. The artificial flattening in this case is much less severe, however please refer to the text for a full discussion of this.}
\label{fig:alphas_sim}  
\end{center}
\end{figure*}

To achieve this we expand on the simulation method used to determine the clean bias in Section \ref{sec:photometry}. Point sources were once again injected into Measurement Sets of 500 of the survey pointings; however, each source was assigned a spectral index value of either 0.0, $-$0.7 or $-$1.4. Spectral curvature in each case was zero. The same twelve logarithmically-spaced flux density bins were retained, but extra iterations were required to build up the statistics at the faint end as sometimes the reconstruction of the synthetic source in the higher order Taylor terms was not satisfactory and no spectral index could be fit. In total approximately 25,000 viable simulations were executed on the Galaxy supercomputer.

Figure \ref{fig:alphas_sim} recreates Figure \ref{fig:alphas} but presents the reconstructed $\alpha$ and $\beta$ measurements for the simulated sources. On the lower two panels the lines coloured blue, green and red track the recovered flux density-dependent spectral indices for the three intrinsic values shown in the inset legend. Large points are the median $\alpha$ and $\beta$ values, again coloured by flux density as indicated at the top of the figure. The black points show the median measurements from the catalogue in each flux density bin, as per Figure \ref{fig:alphas} and Table \ref{tab:medalphas}.

Regarding the $\alpha$ values, the simulations reveal something potentially insidious, in that the apparent flattening of the median spectral index may at least in part be an artefact of the broadband deconvolution approach. A shift towards flatter spectra begins below $\sim$3 mJy, but the rate at which this flattening occurs seems to be a function of the intrinsic steepness of the spectrum. As mentioned in Section \ref{sec:selfcal} a third-order Taylor series was used during imaging. For comparison we run a smaller simulation whereby $\alpha$~=~$-$0.7 sources are simulated for the lowest three flux density bins, but the data are imaged with a second-order Taylor expansion. The results are shown by the crosses on Figure \ref{fig:alphas_sim} (n~=~2 in the legend). Some flattening still occurs but it is less pronounced than in the third-order data. The flux densities at which this flattening begins means that it also cannot be explained via SNR considerations, as our simple two-band simulation illustrates (Figure \ref{fig:two_point_noisy}). This appears to present us with a trade-off: the spectra at the faint end can be recovered more faithfully, at the expense of reducing the dynamic range of the survey images (and therefore increasing the effective noise floor, see Section \ref{sec:sensitivity}). In our limited trial imaging runs increasing the number of terms to 3 significantly improved the quality of the Stokes-I images, particularly around brighter sources. We ascribe this to the significant variation in the extent of the primary beam across the 1--2 GHz of frequency coverage, which imparts pronounced artificial spectral shapes in off-axis sources that are more faithfully deconvolved by the higher-order approach. 

The simulations also confirm that the suspicious behaviour of the observed $\beta$ values is also likely to be an artefact of fitting a higher-order polynomial, with faint sources exhibiting unreasonable levels of curvature in their reconstructed model spectra. For this reason we omit the curvature measurements from the component catalogue. Note that the steepest $\alpha$~=~$-$1.4 simulated sources all have a negative curvature measurement even at the brightest flux densities. While it is reasonable that the $\beta$ measurements derived from the data for the very strongest sources may be accurate, the persistent offset between the measured and simulated curvature values, and the significantly different derivative of the curve is a mystery. We speculate that this may be related to the initial self-calibration step described in Section \ref{sec:selfcal}. A single per-antenna phase-only gain correction is made for each $\sim$2.5 minute snapshot, however the gain corrections are derived by producing a visibility model derived from the components in the Taylor term images, constrained by the FIRST-based cleaning mask. This might give rise to a positive feedback situation whereby anomalous spectral shapes are reinforced by the self-calibration, and then enhanced further in the subsequent round of MT-MFS imaging. This would remain hidden in a flux density comparison (e.g. Figure \ref{fig:photometry}) as MT-MFS was fixing the flux density at the band centre. If this proves to be the case then it suggests that a sub-band self-calibration approach is probably better for such large fractional bandwidths, with independent sky models derived from images formed from smaller fractional bandwidth frequency chunks which can be reliably imaged using 1 or 2 Taylor terms to model the spectra. As with the clean bias we also cannot rule out effects such as survey-wide RFI persisting in certain regions of the spectrum that is not excised by the automatic flagging from playing a part. A spectral window with a flux scale that was skewed high would introduce discontinuities that would affect the reconstructed spectrum.

Unfortunately, although these simulations have proved enlightening in terms of identifying the point where the spectral index measurements may become unreliable, they are unable to provide any heuristic corrections to nullify the biases, cf. the FIRST and NVSS approach to clean bias. The artificial flattening of $\alpha$ with decreasing flux density is clearly dependent on the intrinsic spectral shape of the component being modelled, and since genuine astronomical radio sources will have a range of intrinsic spectral shapes, no single correction factor will be valid.

\begin{table}
\centering
\caption{Median spectral index measurements for components with integrated flux densities between $S_{1}$ and $S_{2}$. The number of components in each bin is denoted by N. These values are plotted as black markers on the lower panel of Figure \ref{fig:alphas}. See the caveats as to the reliability of these measurements at fainter flux densities in Section \ref{sec:alphas}.}
\begin{tabular}{ccccc} \hline
$S$  & $S_{1}$  & $S_{2}$  & $\alpha_{\mathrm{med}}$ & $N$ \\ 
$[$mJy b$^{-1}]$ & $[$mJy b$^{-1}]$ & $[$mJy b$^{-1}]$ &  &  \\ \hline
0.49   & 0.4    & 0.58   & $-$0.37  & 496  \\
0.71   & 0.58   & 0.85   & $-$0.53  & 1176\\
1.04   & 0.85   & 1.23   & $-$0.72  & 1775\\
1.51   & 1.23   & 1.79   & $-$0.78  & 1741\\
2.2    & 1.79   & 2.61   & $-$0.82  & 1446\\
3.21   & 2.61   & 3.8    & $-$0.86  & 1264\\
4.67   & 3.8    & 5.53   & $-$0.88  & 1048\\
6.79   & 5.53   & 8.05   & $-$0.86  & 746\\
9.89   & 8.05   & 11.72  & $-$0.94  & 601\\
14.39  & 11.72  & 17.06  & $-$0.92  & 424\\
20.95  & 17.06  & 24.83  & $-$0.94  & 327\\
30.49  & 24.83  & 36.14  & $-$0.9   & 252\\
44.37  & 36.14  & 52.6   & $-$0.95  & 146\\
64.58  & 52.6   & 76.56  & $-$0.92  & 114\\
93.99  & 76.56  & 111.43 & $-$0.98  & 65\\
136.8  & 111.43 & 162.18 & $-$0.92  & 46\\
199.11 & 162.18 & 236.04 & $-$0.97  & 21\\
289.79 & 236.04 & 343.54 & $-$0.72  & 8\\
421.77 & 343.54 & 500.0  & $-$0.95  & 9\\ \hline
\end{tabular}
\label{tab:medalphas}
\end{table}

\subsection{Eddington and resolution bias corrections, visibility areas, and the differential source counts}
\label{sec:bias}

This section describes measurements of the Euclidean-normalized differential source counts as constructed from the component catalogues, together with estimates of relevant bias corrections. For an overview of the subject of radio source counts we refer the reader to the review by \citet{deZotti2010}. Although our measurements do not break new ground in this area, determining the source counts and associated bias corrections are an additional effective way to validate the results of a radio survey.

The source counts are determined in logarithmically-spaced flux density bins as per Table \ref{tab:counts}. The counts are expressed per unit area, therefore the first correction we determine is the so-called visibility area of each flux density bin in order to compensate for the non-uniform depth of the survey. The visibility area ($A_{vis}$) is simply the sky area over which the source detection procedure will be sensitive to a source with a flux density at or above the lower edge of the bin. This is computed using Figure \ref{fig:rms} by summing the total pixel area with RMS values exceeding one-fifth of the lower bin edge (recalling the 5$\sigma$ peak threshold described in Section \ref{sec:catalogue}). The visibility areas are listed for each bin in Table \ref{tab:counts}. Since the lowest visibility area corresponding to the faintest flux density bin is 40.4 square degrees, the effects of field to field variations on the source counts are negligible \citep{HeywoodJarvisCondon2013}.

\begin{table}
\centering
\caption{Tabulated differential source counts in each flux density bin, centered at $S_{\mathrm{bin}}$, in both raw (N) and Euclidean-normalized form. The uncertainties for the Euclidean-normalized counts are based on the Poisson  ($\sqrt{N}$) uncertainties of the raw counts. Visibility areas ($A_{vis}$) and corrections due to Eddington bias ($f_{\mathrm{E}}$) are also provided.}
\begin{tabular}{cccccc} \hline
$S_{\mathrm{bin}}$   & $A_{vis}$     &   $f_{\mathrm{E}}$ & $N$  & $S^{2.5} dN/dS$          & Uncertainty                    \\
$[$mJy$]$   & $[$sq.deg.$]$ &           &      & $[$Jy$^{1.5}$~sr$^{-1}]$ & $[$Jy$^{1.5}$~sr$^{-1}]$ \\ \hline
0.68   & 40.4  & 1.4 & 334  & 4.0   & 0.1  \\
0.95   & 76.9  & 1.2 & 869  & 5.4   & 0.1  \\
1.34   & 98.1  & 1.0 & 1508 & 7.5   & 0.2  \\
1.89   & 104.3 & 1.0 & 1647 & 10.4  & 0.3  \\ 
2.66   & 105.3 & 1.0 & 1485 & 14.6  & 0.4  \\
3.74   & 105.8 & 1.0 & 1244 & 21.4  & 0.6  \\
5.27   & 105.9 & 1.0 & 1106 & 28.9  & 1.0  \\
7.42   & 105.9 & 1.0 & 892  & 35.9  & 1.4  \\
10.46  & 105.9 & 1.0 & 656  & 49.5  & 2.1  \\
14.73  & 105.9 & 1.0 & 554  & 60.3  & 3.0  \\
20.74  & 105.9 & 1.0 & 391  & 78.5  & 4.5  \\
29.22  & 105.9 & 1.0 & 310  & 103.8 & 6.6  \\
41.15  & 105.9 & 1.0 & 243  & 111.9 & 8.9  \\
57.96  & 105.9 & 1.0 & 159  & 149.4 & 13.3 \\
81.64  & 105.9 & 1.0 & 126  & 144.8 & 16.9 \\ 
114.99 & 105.9 & 1.0 & 73   & 169.0 & 23.7 \\
161.96 & 105.9 & 1.0 & 51   & 193.9 & 32.8 \\
228.11 & 105.9 & 1.0 & 35   & 138.9 & 35.9 \\
321.29 & 105.9 & 1.0 & 15   & 108.3 & 40.9 \\
452.54 & 105.9 & 1.0 & 7    & 207.0 & 73.2 \\
637.39 & 105.9 & 1.0 & 8    & 173.0 & 86.5 \\ \hline
\end{tabular}
\label{tab:counts}
\end{table}

\citet{Eddington1913} bias is a by-product of image noise that causes a redistribution of the source counts in the fainter bins. As the value of a flux density bin approaches the noise level of the image the noise becomes an increasingly significant fraction of the component flux density. Thus there is an increased chance of a random noise spike coinciding with the position of a component and pushing it into a neighbouring bin. Eddington bias is also the reason that it is not advisable to count sources below a threshold of 5$\sigma$, as its effects quickly become devastating below this threshold. Even at 5$\sigma$ one can expect $\sim$6 spurious features in a 10,000~$\times$~10,000 pixel image.
 
Quantifying its effects in the catalogue of a survey such as this, derived using position-dependent thresholds, is not trivial. To estimate the Eddington bias correction for each flux density bin we make use of a simulation of the extragalactic radio source population \citep{Wilman2008,Wilman2010} and the effective survey sensitivity map (Figure \ref{fig:rms}). The 1.4 GHz component flux densities are extracted from a random 100 square degree subset of the simulation. The flux density of each component is then perturbed by an additive noise value drawn from a normal distribution with a mean value of zero, the RMS of which is selected from a weighted distribution based on the survey RMS measurements given in Figure \ref{fig:rms}. This process is repeated a thousand times and the mean `perturbed' source count distribution is measured from the resulting catalogue using the same set of bins as listed in Table \ref{tab:counts}. This process simulates the Eddington bias: the effect of the noise perturbation is to redistribute the counts between bins in a way that becomes increasingly significant with the decreasing SNR of the component. Since we can determine the true source counts from the unperturbed simulation, the factor $f_{\mathrm{E}}$ required to remove the effects of Eddington bias can be estimated by taking the ratio of the true counts to the perturbed counts. This does assume that the model counts are correct, however they agree very well with observed values in the flux density regime we are considering. The correction factors only deviate from unity for the lowest two flux density bins as indicated in Table \ref{tab:counts}.

\begin{figure}
\centering
\includegraphics[width=\columnwidth]{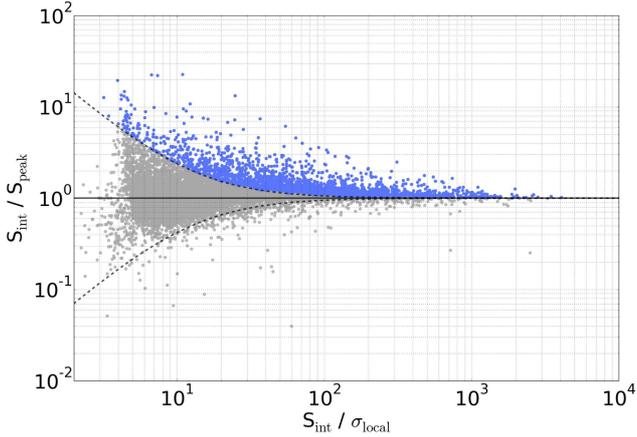}
\caption{The ratio of integrated ($S_{\mathrm{int}}$) to peak ($S_{\mathrm{peak}}$) flux density for each catalogued component as a function of its integrated flux density expressed in units of the local RMS noise ($\sigma$). In the high signal to noise regime the ratio $S_{\mathrm{int}}$~/~$S_{\mathrm{peak}}$ being greater than 1 is a robust indicator that a component is spatially resolved. The dashed lines indicate the `reliability locus' above which a component is deemed to be resolved (the blue points).}
\label{fig:resolved}
\end{figure}

The final effect we investigate is the resolution bias (distinct from the resolution mismatch discussed in Section \ref{sec:photometry}). Resolution bias is defined here as a underestimation of source counts in a particular flux density bin due to a resolved component having a lower peak flux density than an unresolved component with equivalent total (integrated) flux density. Examination of the effect in other radio surveys suggests that at the 16 arcsec~$\times$~10 arcsec resolution of our images the resulting catalogue will not be significantly affected. However we verify this via the analytic method used by e.g.~\citet{Prandoni2001} and \citet{Huynh2005}. 

If the ratio of the integrated ($S_{\mathrm{int}}$) to peak ($S_{\mathrm{peak}}$) flux density of a detected component is greater than unity then this is a robust indicator that at the component is spatially resolved, with
\begin{equation}
\frac{S_{\mathrm{int}}}{S_{\mathrm{peak}}}~=~\frac{\theta_{\mathrm{maj}}\theta_{\mathrm{min}}}{b_{\mathrm{maj}}b_{\mathrm{min}}}
\label{eq:s_int}
\end{equation}
where $\theta_{\mathrm{maj}}$ and $\theta_{\mathrm{min}}$ are the (deconvolved) major and minor axes of the component and $b_{\mathrm{maj}}$ and $b_{\mathrm{min}}$ are the major and minor axis of the restoring beam. The reliability of this metric depends on the SNR ($S_{\mathrm{int}}$~/~$\sigma$) of the detection, i.e.~a source detected at low significance would require $S_{\mathrm{int}}$~/~$S_{\mathrm{peak}}$~$\gg$~1. The first step is to determine a `reliability envelope' on a $S_{\mathrm{int}}$~/~$S_{\mathrm{peak}}$ vs SNR plot, as illustrated on Figure \ref{fig:resolved}. Components below the horizontal line corresponding to an integrated-to-peak flux density ratio of 1 exhibit scatter that increases with decreasing SNR. These components are taken to indicate the region on the plot for which the integrated-to-peak flux density ratio is an unreliable indicator that a component is resolved as it may be dominated by noise in the measurement. Reasoning that there is an equivalent distribution of sources above the unity line, a function below the line is defined that encompasses 90 percent of the points, as indicated by the dashed line. This boundary is mirrored above the plot, represented by the function
\begin{equation}
\frac{S_{\mathrm{int}}}{S_{\mathrm{peak}}}~=~1~+~\left[\frac{c}{\mathrm{SNR}^{1.4}}\right]
\label{eq:resolved}
\end{equation}
where $c$ is determined by iteration to be 35. Using Equation \ref{eq:s_int} it follows that for a survey with a peak flux density limit of 5$\sigma$
\begin{equation}
\frac{S_{\mathrm{int}}}{5\sigma}~=~\frac{\theta_{\mathrm{max}}^{2}}{b_{\mathrm{maj}}b_{\mathrm{min}}},
\end{equation}
i.e.~for a given $S_{\mathrm{int}}$ the maximum angular-size that a source can have before its peak flux density drops below the detection threshold is
\begin{equation}
\theta_{\mathrm{max}}~=~\left[\frac{S_{\mathrm{int}}}{5\sigma}(b_{\mathrm{maj}}b_{\mathrm{min}})\right]^{0.5}.
\label{eq:thetamax}
\end{equation}
Conversely the envelope defined by Equation \ref{eq:resolved} can be combined with Equation \ref{eq:s_int} to give
\begin{equation}
\theta_{\mathrm{min}}~=~\left[(b_{\mathrm{maj}}b_{\mathrm{min}})\left(1~+~\left[\frac{c}{\mathrm{SNR}^{1.4}}\right]\right)\right]^{0.5}
\label{eq:thetamin}
\end{equation}
where $\theta_{\mathrm{min}}$ is the minimum angular size a source can have before it can be reliably deemed to be resolved as a function of its signal to noise ratio. Using the two angular size limits defined by Equations \ref{eq:thetamax} and \ref{eq:thetamin} together with an assumed true source angular size distribution allows estimation of the incompleteness corrections required for the data due to component resolution biases. Following \citet{Windhorst1990} the true size distribution $h(\theta)$ is assumed to follow the exponential function
\begin{equation}
h(\theta)~=~\mathrm{exp}\left[-\mathrm{ln}~2\left(\frac{\theta_{\mathrm{lim}}}{\theta_{\mathrm{med}}}\right)^{0.62}\right]
\end{equation}
where
\begin{equation}
\theta_{\mathrm{lim}}~=~\mathrm{max}(\theta_{\mathrm{min}},\theta_{\mathrm{max}})
\end{equation}
and
\begin{equation}
\theta_{\mathrm{med}}~=~2S^{0.3}
\end{equation}
where S is the flux density in mJy. The correction factor $f_{\mathrm{R}}$ for the counts is then given by
\begin{equation}
f_{\mathrm{R}}~=~\frac{1}{1 - h(\theta)}.
\end{equation}
In all the above calculations an area-weighted mean value of $\sigma$ is used. The resulting resolution bias correction factors are unity for all flux density bins. As expected, the counts do not need to be adjusted to compensate for this effect.

\begin{figure}
\centering
\includegraphics[width=\columnwidth]{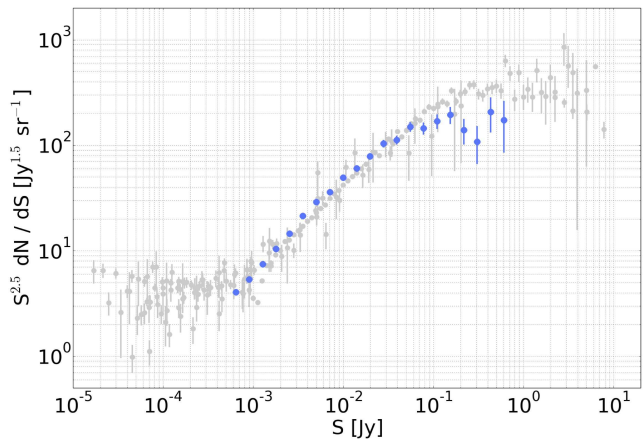}
\caption{Euclidean-normalized differential source counts derived from the Stripe 82 survey (blue points) overlaid on other observationally-derived measurements by various authors, as tabulated by \citet{deZotti2010}.}
\label{fig:source_counts}
\end{figure}

Figure \ref{fig:source_counts} shows the final values of the differential source counts, normalised to a Euclidean universe, with the visibility area and Eddington bias corrections applied. The raw and normalised counts, along with the relevant correction parameters are listed in Table \ref{tab:counts}. The values derived agree with existing observations. The significant discrepancies at the bright end are due to the relatively small area of the survey in terms of studying these rare and bright objects. We have made no attempts to compensate for this incompleteness.

\section{Conclusions}
\label{sec:conclusions}

We have used the Karl G.~Jansky Very Large Array to conduct a snapshot survey covering $\sim$100 square degrees of the SDSS Stripe 82 region, reaching a thermal noise level of approximately 50~$\mu$Jy beam$^{-1}$, but with a typical 1$\sigma$ effective depth of 88~$\mu$Jy beam$^{-1}$. The use of the CnB configuration results in a survey with good sensitivity to low surface brightness and diffuse emission, highly synergistic with existing 1.4 GHz radio imaging of the field. Our component catalogue contains 11,782 point and Gaussian components, associated with approximately 8,948 unique radio sources. Photometric and astrometric performance of the survey is excellent, as verified by comparison with measurements from three external radio surveys of varying resolution and depth covering the same field. The differential source counts derived from the catalogue are in good agreement with previous studies.

We used the full 1--2 GHz bandwidth available with the VLA's L-band system, and in doing so have produced one of the first large area radio surveys to feature in-band spectral index estimates in the catalogue without having to cross match the results with an external survey at a second frequency. Spectral measurements were derived by forming images from the entire bandwidth using the MT-MFS algorithm, identifying components in the Stokes-I images and then extracting error-weighted measurements of $\alpha$ and $\beta$ for each component. The clean bias effect remains an issue for snapshot surveys with the VLA, despite our best efforts to minimise it with a conservative approach to deconvolution. However it also appears that there are significant biases in the spectral domain associated with the MT-MFS algorithm. The apparent trend towards flatter spectra with decreasing flux density is shown to be at least partially artificial, and dependent on the intrinsic spectral shape of the source, and the behaviour of the spectral curvature estimates is not deemed to be reliable, although this is expected for anything but the brightest components. Above a threshold of $\sim$3~mJy we deem our median $\alpha$ measurements to be reliable, and in good agreement with existing measurements derived by cross-matching the results of pairs of independent observing programs at different frequencies.

The use of large scale simulations to investigate the biases in the imaging process has proven to be very informative, however unlike surveys such as FIRST and NVSS it does not seem practical to derive a single correction factor for the clean bias, which appears to vary significantly from pointing to pointing. A significant contributing factor is likely to be the changes in the PSF shape that arise due to the dynamic scheduling of the observations, which impart an undesirable level of non-uniformity to the survey. A deeper examination of how the clean bias correlates with factors such as the PSF shape and the level of residual RFI will be pursued. Although single correction factors to the Stokes-I flux density and the spectral measurements cannot be determined by running large-scale simulations on a per-observation basis, as was the method for FIRST and NVSS, an estimate of the overall level of the biases can be obtained, and indeed should be, in order to properly understand the limitations of survey products. The simulations were computationally expensive and required the use of a supercomputer, although we note that this may be a natural application for cloud computing platforms.	

Finally, the vastly increased survey potential of the now complete VLA upgrade cannot be understated. The sub-millijansky radio sky can now be probed with a snapshot survey with very modest integration times per pointing, and the potential for the instantaneous characterisation of the bulk spectral behaviour of radio sources without the need for conducting two surveys (with mismatched characteristics) is huge, provided the associated systematics, examples of which are highlighted in this paper, can be understood and mitigated. 

Our images and catalogues are publicly available, and can be retrieved from the Australia Telescope National Facility's FTP server: {\tt \href{fftp://ftp.atnf.csiro.au/pub/people/hey036/Stripe82}{ftp://ftp.atnf.csiro.au/pub/people/hey036/Stripe82}}.

\section*{Acknowledgements}
\addcontentsline{toc}{section}{Acknowledgements}

We thank the anonymous referee and the MNRAS editorial staff for providing very useful comments on this paper. The National Radio Astronomy Observatory is a facility of the National Science Foundation operated under cooperative agreement by Associated Universities, Inc. This work was supported by resources provided by the Pawsey Supercomputing Centre with funding from the Australian Government and the Government of Western Australia. IH thanks the Rhodes Centre for Radio Astronomy Techniques and Technologies (RATT) for the provision of computing facilities. This research has made use of NASA's Astrophysics Data System. This research made use of Montage. It is funded by the National Science Foundation under Grant Number ACI-1440620, and was previously funded by the National Aeronautics and Space Administration's Earth Science Technology Office, Computation Technologies Project, under Cooperative Agreement Number NCC5-626 between NASA and the California Institute of Technology. Some figures in this paper were created using the Python package APLpy, an open-source plotting package for Python hosted at {\tt \href{http://aplpy.github.com}{http://aplpy.github.com}}. IH acknowledges useful discussions with Natasha Maddox. IH thanks the participants of the SAGE workshop, and SKA South Africa for their hospitality during this event.



 
%





\bsp	
\label{lastpage}
\end{document}